\input harvmac
\input epsf
\input labeldefs.tmp
\writedefs

\def\ie{{\it i.e.\/}}
\def\bfone{\relax{\rm 1\kern-.35em 1}}
\def\inbar{\vrule height1.5ex width.4pt depth0pt}

\def\IC{\relax\,\hbox{$\inbar\kern-.3em{\rm C}$}}
\def\ID{\relax{\rm I\kern-.18em D}}
\def\IF{\relax{\rm I\kern-.18em F}}
\def\IH{\relax{\rm I\kern-.18em H}}
\def\II{\relax{\rm I\kern-.17em I}}
\def\IN{\relax{\rm I\kern-.18em N}}
\def\IP{\relax{\rm I\kern-.18em P}}
\def\IQ{\relax\,\hbox{$\inbar\kern-.3em{\rm Q}$}}
\def\us#1{\underline{#1}}
\def\IR{\relax{\rm I\kern-.18em R}}
\font\cmss=cmss10 \font\cmsss=cmss10 at 7pt
\def\ZZ{\relax\ifmmode\mathchoice
{\hbox{\cmss Z\kern-.4em Z}}{\hbox{\cmss Z\kern-.4em Z}}
{\lower.9pt\hbox{\cmsss Z\kern-.4em Z}}
{\lower1.2pt\hbox{\cmsss Z\kern-.4em Z}}\else{\cmss Z\kern-.4em
Z}\fi}

\def\cN{{\cal N}} 
 
 \def\cS{{\cal S}}
\def\cU{{\cal U}} 
\def\nup#1({Nucl.\ Phys.\ $\us {B#1}$\ (}
\def\plt#1({Phys.\ Lett.\ $\us  {B#1}$\ (}
\def\cmp#1({Comm.\ Math.\ Phys.\ $\us  {#1}$\ (}
\def\prp#1({Phys.\ Rep.\ $\us  {#1}$\ (}
\def\prl#1({Phys.\ Rev.\ Lett.\ $\us  {#1}$\ (}
\def\prv#1({Phys.\ Rev.\ $\us  {#1}$\ (}
\def\mpl#1({Mod.\ Phys.\ Let.\ $\us  {A#1}$\ (}
\def\ijmp#1({Int.\ J.\ Mod.\ Phys.\ $\us{A#1}$\ (}
\def\jag#1({Jour.\ Alg.\ Geom.\ $\us {#1}$\ (}
\def\tit#1|{{\it #1},\ }

\def\Coe#1.#2.{{#1\over #2}}
\def\coeff#1#2{\relax{\textstyle {#1 \over #2}}\displaystyle}
\def\coe#1.#2.{\relax{\textstyle {#1 \over #2}}\displaystyle}
\def\half{{1 \over 2}}

\def\del{\partial}

%

\def\TR{$\displaystyle{{}##}$\hfil}
\def\TC{\hfil$\displaystyle{##}$\hfil}
\def\TT{\hbox{##}}
\def\seqalign#1#2{\vcenter{\openup1\jot
  \halign{\strut #1\cr #2 \cr}}}


\def\comment#1{}
\def\fixit#1{}

\def\tf#1#2{{\textstyle{#1 \over #2}}}

\def\mop#1{\mathop{\rm #1}\nolimits}


\def\diag{\mop{diag}}
\def\tr{\mop{tr}}


\def\lsim{\mathrel{\mathstrut\smash{\ooalign{\raise2.5pt\hbox{$<$}
\cr\lower2.5pt\hbox{$\sim$}}}}}
\def\gsim{\mathrel{\mathstrut\smash{\ooalign{\raise2.5pt\hbox{$>$}
\cr\lower2.5pt\hbox{$\sim$}}}}}



\def\sqr#1#2{{\vcenter{\vbox{\hrule height.#2pt
         \hbox{\vrule width.#2pt height#1pt \kern#1pt
            \vrule width.#2pt}
         \hrule height.#2pt}}}}




\def\href#1#2{#2}


\def\TR{$\displaystyle{{}##}$\hfil}
\def\TC{\hfil$\displaystyle{##}$\hfil}
\def\TT{\hbox{##}}
\def\seqalign#1#2{\vcenter{\openup1\jot
  \halign{\strut #1\cr #2 \cr}}}
\def\r#1#2{{\bf #1}_{#2}}

\def\La{\Lambda}
\def\eql{{~=~}}
\def\tO{{\widetilde O}}
\def\Om{{\Omega}}
\def\Si{{\Sigma}}
\def\al{\alpha}
\def\be{\beta}
\def\Ga{{\Gamma}}
\def\tr{{\rm Tr\,}}

\def\calV{{\cal V}}
\def\ggamma{{\tilde \Gamma}}

\def\nn#1#2{n_{#1}^{(#2)}}


%
\def\nihil#1{{\it #1}}
\def\eprt#1{{\tt #1}}

\lref\GRWplb{M.\ G\"unaydin, L.J.\ Romans and N.P.\ Warner,
\nihil{Gauged $N=8$ Supergravity in Five Dimensions,}
Phys.~Lett.~{\bf 154B} (1985) 268.}
\lref\GRW{M.\ G\"unaydin, L.J.\ Romans and N.P.\ Warner, \nihil{
Compact and Non-Compact
Gauged Supergravity Theories in Five Dimensions,}
Nucl.~Phys.~{\bf B272} (1986) 598.}
\lref\KPW{A.~Khavaev, K.~Pilch and N.P.~Warner,
\nihil{New vacua of gauged N=8 supergravity in five-dimensions,}
\eprt{hep-th/9812035}.}
\lref\FFZ{S.\ Ferrara, C.\ Fronsdal and A.\ Zaffaroni,
 \nihil{On N=8 supergravity on AdS(5) and N=4 superconformal
Yang--Mills
theory,} Nucl.~Phys.~{\bf B532} (1998) 153, \eprt{hep-th/9802203}.}
\lref\SupCurr{E.\ Bergshoeff, M.\ de Roo and B.\ de Wit,
\nihil{Extended
Conformal  Supergravity,} Nucl. Phys. {\bf B182} (1981) 173; \hfill
\break
P.\ Howe, K.S.\ Stelle, P.K.\ Townsend,  \nihil{Supercurrents,}
Nucl.~Phys.~{\bf B192} (1981) 332.}
\lref\GPPZ{L.\ Girardello, M.\ Petrini, M.\ Porrati and
A.\ Zaffaroni,  \nihil{Novel Local CFT and Exact Results on
Perturbations
of $N=4$ Super Yang--Mills from AdS Dynamics,} JHEP {\bf 12} (1998) 022,
\eprt{hep-th/9810126}.  See {\tt hep-th} version~4 or later, or the
version published in JHEP, for discussion of the c-function.}
\lref\JDFZ{J.\ Distler and F.\ Zamora,
\nihil{Nonsupersymmetric Conformal Field Theories from Stable Anti-de
Sitter Spaces,}  \eprt{hep-th/9810206}.}
\lref\PBDF{P.\ Breitenlohner and D.Z.\ Freedman,
\nihil{Positive
Energy in anti-de Sitter Backgrounds and Gauged Extended Supergravity,}
Phys.~ Lett.~{\bf 115B} (1982) 197; \nihil{Stability  in Gauged
Extended
Supergravity,}  Ann.~Phys.~{\bf 144} (1982) 249.}
\lref\MezTow{
L.~Mezincescu and P.K.~Townsend,
{\it Stability at a Local Maximum in Higher Dimensional anti-de Sitter
Space and Applications to Supergravity},
Ann. Phys. {\bf 160}  (1985) 406.
}
\lref\PosEn{L.\ F.\ Abbott and S.\ Deser,  \nihil{Stabilty of Gravity
with a Cosmological Constant,}  Nucl.~Phys.~{\bf B195} (1982) 76;
\hfil\break
G.W.\ Gibbons, C.M.\ Hull and N.\ P.\ Warner,  \nihil{The Stability
of Gauged Supergravity}  Nucl.~Phys.~{\bf B218} (1983) 173;
\hfil\break
W.\ Boucher, \nihil{Positive Energy without Supersymmetry,}
Nucl.~Phys.~{\bf B242} (1984) 282; \hfil \break
L.\ Mezincescu and P.K.\ Townsend, \nihil{Positive Energy and the
Scalar Potential in Higher Dimensional (Super)Gravity Theories,}
Phys.~Lett.~{\bf 148B} (1984) 55; \hfil \break
L.\ Mezincescu and P.K.\ Townsend, \nihil{Stability  at a Local
Maximum in Higher Dimensional Anti-de Sitter Space and Applications
to Supergravity,} Ann.~Phys.~{\bf 160} (1985) 406.}
\lref\SGIKAP{S.S.\ Gubser, I.R.\ Klebanov, A.M.\ Polyakov,
\nihil{Gauge Theory  Correlators from Non-Critical String Theory,}
Phys.~Lett.~{\bf B428}  (1998) 105, \eprt{hep-th/9802109}.}
\lref\WitHolOne{E.\ Witten, \nihil{Anti-de Sitter space and
holography,} Adv. Theor.  Math.  Phys. {\bf 2} (1998) 253,
\eprt{hep-th/9802150}.}
\lref\SSG{S.S.\ Gubser, \nihil{Einstein Manifolds and Conformal
Field Theories,}  \eprt{hep-th/9807164}}
\lref\MHKS{M.\ Henningson, K.\ Skenderis,  \nihil{The Holographic Weyl
Anomaly,}  J.~High Energy Phys.~{\bf 9807} (1998) 023,
\eprt{hep-th/9806087}.}
\lref\OPEfourD{D.\ Anselmi, M.\ Grisaru and A.\ Johansen,
\nihil{A Critical Behavior of Anomalous Currents, Electric--Magnetic
Universality and CFT in Four Dimensions,} Nucl.~Phys.~{\bf B491}
(1997) 221-248, \eprt{hep-th/9601023};  \hfil \break
D.\ Anselmi, D.\ Z.\ Freedman, M.\ T.\ Grisaru, A.\ A.\ Johansen,
\nihil{Universality of the Operator Product Expansions of SCFT
in Four Dimensions,} Phys.~Lett.~{\bf B394} (1997) 329-336,
\eprt{hep-th/9608125};  \hfil \break
D.\ Anselmi, \nihil{The N=4 Quantum Conformal Algebra,}
\eprt{hep-th/9809192}; \nihil{Quantum Conformal Algebras and Closed
Conformal Field Theory,} \eprt{hep-th/9811149}.}
\lref\BdWHN{B.\ de Wit, H.\ Nicolai, \nihil{The Consistency of the
$S^7$  Truncation in $d = 11$ Supergravity,} Nucl.~Phys.~{\bf B281}
(1987) 211.}
\lref\IRKEW{I.R.\ Klebanov and E. Witten, \nihil{Superconformal
Field Theory on Three-Branes at a Calabi-Yau Singularity,}
\eprt{hep-th/9807080}.}
\lref\samson{S.~S.~Gubser, N.~Nekrasov, and S.~Shatashvili,
\nihil{Generalized conifolds and 4-dimensional $N=1$ superconformal field
theory,} \eprt{hep-th/9811230}.}
\lref\gkSchwing{S.~S.~Gubser and I.~R.~Klebanov, \nihil{Absorption by branes and
Schwinger terms in the world-volume theory,} Phys.~Lett.~{\bf B413} (1997)
41.}
\lref\lopez{E.~Lopez, \nihil{A family of $N=1$ $SU(N)^k$ theories from
branes at singularities,} { JHEP} {\bf 2}
(1998) 019, \eprt{hep-th/9812025}.}
\lref\SKES{S.\ Kachru and E.\ Silverstein,
\nihil{4-D Conformal Theories and Strings on Orbifolds,}
Phys. Rev. Lett. {\bf 80} (1998) 4855, \eprt{hep-th/9802183}.}
\lref\JMalda{J.~Maldacena, \nihil{The Large $N$ Limit of Superconformal
Field Theories and Supergravity,}, Adv.~Theor. Math. Phys.~{\bf 2}
(1998) 231 \eprt{hep-th/9711200}.}
\lref\MAPT{M.~ Awada and P.K.~Townsend,
 \nihil{$N=4$ Maxwell Einstein Supergravity in Five Dimensions
and its $SU(2)$ Gauging,} Nucl.~Phys.~{\bf B255} (1985) 617 .}
\lref\ECrem{E.~ Cremmer,
\nihil{Supergravities in Five Dimensions,} in {\it
Superspace and Supergravity,} S.W.~Hawking and M.~Ro\v cek  (editors),
Proceedings of the Nuffield Gravity Workshop, Cambridge,
Jun 16 -- Jul 12, 1980;  Cambridge  Univ. Pr. (1981) .}
\lref\DOEW{E.~Witten, D.~Olive, \nihil{Supersymmetry Algebras that
include Topological Charges,} Phys.~ Lett.~{\bf 78B} (1978) 97.}
\lref\LNV{A.~Lawrence, N.~Nekrasov and C.~Vafa, \nihil{On conformal field
theories in four-dimensions,} Nucl.~ Phys.~{\bf B533} (1998) 199,
\eprt{hep-th/9803015}.}
\lref\AMYH{A.~ Hanany and Yang-Hui He, \nihil{NonAbelian finite
gauge theories,} \eprt{hep-th/9811183}.}
\lref\Zamo{A.\ B.\ Zamolodchikov,
 \nihil{``Irreversibility'' of the Flux of the Renormalization Group
in a 2-d Field Theory,} JETP Lett.~{\bf 43} (1986) 730-732.}
\lref\NSVZ{
V.~Novikov, M.~A.\ Shifman, A.~I.\ Vainshtein, V.~Zakharov, {\it Exact
Gell-Mann-Low Function of Supersymmetric Yang-Mills Theories from
Instanton
Calculus,} Nucl. Phys. {\bf B229} (1983) 381.}
\lref\ksv{
I.~I.\ Kogan, M.~A.\ Shifman, and A.~I.\ Vainshtein, {\it Matching
Conditions and Duality in N=1 SUSY Gauge Theories in the Conformal
Window,} Phys. Rev.~{\bf D53} (1996) 4526, \eprt{hep-th/9507170}.} 
\lref\Appelquist{T.~Appelquist, A.~G.~Cohen, and M.~Schmaltz, 
\nihil{A new constraint on strongly coupled gauge theories,} 
\eprt{hep-th/9901109}.}
\lref\kw{
I.~R. Klebanov and E.~Witten, {\it Superconformal field theory on
three-branes
  at a Calabi-Yau singularity,} { Nucl. Phys.} {\bf B536} (1998) 199,
  {{\tt hep-th/9807080}}.}
\lref\abks{ O.~Aharony, M.~Berkooz, D.~Kutasov, and N.~Seiberg, {\it
Linear dilatons, NS five-branes and holography,} { JHEP} {\bf 10}
(1998) 004, {{\tt hep-th/9808149}}.}
\lref\WitHolTwo{ E.~Witten, {\it Anti-de Sitter space, thermal phase
transition, and confinement in gauge theories,} { Adv. Theor. Math.
Phys.} {\bf 2} (1998) 505, {{\tt hep-th/9803131}}.}

\lref\ktZero{ I.~R. Klebanov and A.~A. Tseytlin, {\it Asymptotic
freedom and infrared behavior in the type 0 string approach to gauge
theory,} {{\tt hep-th/9812089}}.}

\lref\minConfine{ J.~A. Minahan, {\it Asymptotic freedom and
confinement from type 0 string theory,} {{\tt hep-th/9902074}}.}
\lref\gDil{ S.~S. Gubser, {\it Dilaton driven confinement,} {{\tt
hep-th/9902155}}.}
\lref\imsy{ N.~Itzhaki, J.~M. Maldacena, J.~Sonnenschein, and
S.~Yankielowicz, {\it Supergravity and the large N limit of theories
with sixteen supercharges,} { Phys. Rev.} {\bf D58} (1998) 046004,
\href{<<<http://xxx.lanl.gov/abs/hep-th/9802042}{{\tt
hep-th/9802042}}.}
\lref\ls{
R.~G. Leigh and M.~J. Strassler, {\it Exactly marginal operators and
duality in four-dimensional N=1 supersymmetric gauge theory,} { Nucl.
Phys.} {\bf B447} (1995) 95--136,
\href{<<<http://xxx.lanl.gov/abs/hep-th/9503121}{{\tt
hep-th/9503121}}.}
\lref\isReview{ K.~Intriligator and N.~Seiberg, {\it Lectures on
supersymmetric gauge theories and electric - magnetic duality,} { Nucl.
Phys. Proc. Suppl.} {\bf 45BC} (1996) 1--28,
\href{<<<http://xxx.lanl.gov/abs/hep-th/9509066}{{\tt
hep-th/9509066}}.}

\lref\BanksZaks{
T.~Banks and A.~Zaks, {\it On the Phase Structure of Vector--like
Gauge Theories with Massless Fermions,} { Nucl. Phys.} {\bf B196}
(1982) 189.}
\lref\klm{ A.~Karch, D.~L\"ust, and A.~Miemiec, {\it New N=1
superconformal field theories and their supergravity description,}
\href{<<<http://xxx.lanl.gov/abs/hep-th/9901041}{{\tt
hep-th/9901041}}.}
\lref\StrasslerPrivate{
M. Strassler, private communication.}
\lref\afgj{ D.~Anselmi, D.~Z. Freedman, M.~T. Grisaru, and A.~A.
Johansen, {\it Nonperturbative formulas for central functions of
supersymmetric gauge theories,} { Nucl. Phys.} {\bf B526} (1998) 543,
{{\tt hep-th/9708042}}.}
\lref\gEin{ S.~S. Gubser, {\it Einstein manifolds and conformal field
theories,} { Phys. Rev.} {\bf D59} (1999) 025006, {{\tt
hep-th/9807164}}.}
\lref\WitSuss{ L.~Susskind and E.~Witten, {\it The Holographic bound in
anti-de Sitter space,} {{\tt hep-th/9805114}}.}

\lref\PeetPolch{ A.~W. Peet and J.~Polchinski, {\it UV/IR relations
in AdS dynamics,} {\tt hep-th/9809022}.}
\lref\MDGM{M.R.\ Douglas and G.\ Moore,
\nihil{D-branes, Quivers, and ALE Instantons,} 
\eprt{hep-th\slash9603167}.}
\lref\Romans{L.J.\ Romans,
\nihil{New Compactifications of Chiral N=2 D = 10 Supergravity,}
 Phys.\ Lett.\ {\bf 153B} 392 (1985).}
\lref\LJRGgs{L.J.\ Romans,  \nihil{Gauged N=4 Supergravities
in Five-Dimensions and Their Magnetovac Backgrounds,}
Nucl.\ Phys.\ {\bf B267} 433 (1986).}
\lref\PvNNW{P.\ van Nieuwenhuizen and N.P.\ Warner,
\nihil{New Compactifications of ten-dimensional and
eleven-dimensional supergravity on manifolds which are not direct
products,} Commun.\ Math.\ Phys.\ {\bf 99} 141 (1985).}
\lref\GEASAF{G.E.\ Arutyunov and S.A.\ Frolov,
\nihil{Antisymmetric tensor field on $AdS^5$,}
 Phys.\ Lett.\ {\bf B441} 173 (1998),
\eprt{hep-th/9807046}.}
\lref\PvNT{P.K.~Townsend, K.~Pilch and P.~van Nieuwenhuizen,
{\it Selfduality in odd dimensions,} Phys. Lett. {\bf 136B} (1984) 38.}
\lref\lYi{W.~S.~l'Yi, \nihil{Correlators of currents corresponding to the
massive p form fields in AdS/CFT correspondence,} Phys.~Lett.~{\bf B448}
(1999) 218, \eprt{hep-th/9811097}.}
%

\lref\Shuster{
E. Shuster, {\it Killing spinors and Supersymmetry on AdS},
{\tt hep-th/9902129}.}

\lref\apty{O.~Aharony, J.~Pawe\l czyk, S.~Theisen, and
S.~Yankielowicz, \nihil{A note on anomalies in the AdS/CFT
correspondence}, {\tt hep-th/9901134}.}

\lref\AnselmiKehagias{D.~Anselmi and A.~Kehagias, \nihil{Subleading
corrections and central charges in the AdS/CFT correspondence,}
{\tt hep-th/9812092}.}

\lref\ForteLatorre{S.~Forte and J.~I.~Latorre, \nihil{A proof of the
irreversibility of renormalization group flows in four dimensions,} 
Nucl.~Phys.~{\bf B535} (1998) 709, \eprt{hep-th/9805015}.}
\lref\AnsAPrime{D.~Anselmi, \nihil{Anomalies, Unitarity and Quantum
Irreversibility,} \eprt{hep-th/9903059}.}
\lref\AlvarezGomez{E.~Alvarez and C.~Gomez, \nihil{Geometric Holography,
the Renormalization Group and the c-Theorem}, Nucl.~Phys.~{\bf B541} (1999)
441, \eprt{hep-th/9807226}.} 

\lref\Wald{R.~M.~Wald, {\it General Relativity}.  Chicago University Press,
Chicago (1984).}

\lref\SfetsosKehagias{A.~Kehagias and K.~Sfetsos, \nihil{On running
couplings in gauge theories from type IIB supergravity,} 
{\tt hep-th/9902125}.}

\lref\PorratiConfine{L.~Girardello, M.~Petrini, M.~Porrati, A.~Zaffaroni,
\nihil{Confinement and condensates without fine tuning in supergravity
duals of gauge theories,} {\tt hep-th/9903026}.}

\lref\PPvN{
M.~Pernici, K.~Pilch and P.~van Nieuwenhuizen,
{\it Gauged N=8 D=5 Supergravity},
Nucl. Phys. {\bf B259} (1985) 460.}
\lref\Helg{
S. Helgason, {\it Differential Geometry and Symmetric Spaces,}
Academic Press, New York (1962).}
\lref\WessZumino{
J.\ Wess and B.\ Zumino, {Nucl. Phys} {\bf B77} (1974) 73.
}
\lref\FerraraZumino{
S. Ferrara and B. Zumino, {\it Supergauge invariant Yang-Mills
theories},  Nucl. Phys. {\bf B79} (1974) 413.}
\lref\Sohnius{
M.F.~Sohnius, {\it Introducing Supersymmetry,}
Phys. Rep. {\bf 128} (1985) 39. }
\lref\FlatFrons{M.~Flato and C.~Fronsdal,
{\it Representations of Conformal Supersymmetry,}
Lett. Math. Phys. {\bf 8} (1984)  159.}
\lref\DobrPetk{
V.K.~Dobrev and V.B.~Petkova, {\it All Positive Energy Unitary
Irreducible Representations of Extended Conformal Supersymmetry},
Phys. Lett. {\bf 162B} (1985) 127.}
\lref\fmmr{
D.~Z.~Freedman, S.~D.~Mathur, A.~Matusis, and L.~Rastelli,
\nihil{Correlation functions in the $\hbox{CFT}_D/AdS_{D+1}$
correspondence}, \eprt{hep-th/9804058}.}
\lref\FerrZaff{
S.~Ferrara and A.~Zaffaroni,
{\it N=1, N=2 4-D superconformal field theories and supergravity in
$AdS_5$,}
Phys. Lett. {\bf B431} (1998) 49, {\tt
hep-th/9803060}.}
\lref\FerrLLZaff{
S.~Ferrara, M.A.~Lledo and A.~Zaffaroni, {\it Born-Infeld corrections
to D3-brane action in $AdS_5\times S_5$ and N=4, d=4 primary
superfields}, Phys. Rev. {\bf D58} (1998) 105029, {\tt
hep-th/9805082}.}
\lref\KacWak{
V.~Kac and M.~Wakimoto, {\it Integrable highest weight modules over
affine superalgebras and number theory}, in {\it Lie Theory and
Geometry},
J. Brylinski et al. (eds.), Birkhauser, Boston (1994).}
\lref\HSspin{
M.~Henningson and K.~Sfetsos, \nihil{Spinors and the AdS/CFT
correspondence,} Phys.\ Lett.\ {\bf B431} (1998) 63, \eprt{hep-th/9803251}.}
\lref\AVol{A.~Volovich,
{\it Rarita-Schwinger field in the AdS/CFT correspondence},
JHEP {\bf 09} (1998) 022, {\tt hep-th/9809009}.}
\lref\KosRyt{ A.S.~Koshelev and O.A.~Rytchkov, {\it Note on the massive
Rarita-Schwinger field in the AdS/CFT correspondence}, {\tt
hep-th/9812238}.}

%
\parskip=4pt plus 15pt minus 1pt
\baselineskip=12pt plus 2pt minus 1pt
\Title{\vbox{
\hbox{CERN-TH/99-86, \  HUTP-99/A015  }
\hbox{USC-99/1, \ MIT-CTP-2846}
\hbox{{\tt hep-th/9904017}}
}}
{\vbox{\vskip -1.0cm
\centerline{\hbox{Renormalization Group Flows from Holography--}}
\vskip 8 pt
\centerline{\hbox{Supersymmetry and a c-Theorem }}}}
\vskip -.3cm
\centerline{D.~Z.~Freedman$^1$, S.~S.~Gubser$^2$, K.~Pilch$^3$ and 
N.~P.~Warner$^4$\footnote{*}{\rm On
leave from Department of Physics and Astronomy,
USC,  Los Angeles, CA 90089-0484} }
\bigskip
\centerline{$^1${\it Department of Mathematics and
Center for Theoretical Physics,}}
\centerline{{\it Massachusetts Institute of Technology,
Cambridge, MA  02139}}

\smallskip
\centerline{$^2${\it Department of Physics, Harvard University,
Cambridge,
MA  02138, USA}}
\smallskip
\centerline{$^3${\it Department of Physics and Astronomy,
University of Southern California,}}
\centerline{{\it Los Angeles, CA 90089-0484, USA}}
\smallskip
\centerline{$^4${\it Theory Division, CERN, CH-1211 Geneva 23,
Switzerland}}

\bigskip

\centerline{{Abstract}}
\medskip

We obtain first order equations that determine a supersymmetric kink
solution in five-dimensional ${\cal N}=8$ gauged supergravity. The kink
interpolates between an exterior anti-de Sitter region with maximal
supersymmetry and an interior anti-de Sitter region with one quarter of the
maximal supersymmetry. One eighth of supersymmetry is preserved by the kink
as a whole. We interpret it as describing the renormalization group flow in
${\cal N}=4$ super-Yang-Mills theory broken to an ${\cal N}=1$ theory by
the addition of a mass term for one of the three adjoint chiral
superfields.  A detailed correspondence is obtained between fields of bulk
supergravity in the interior anti-de Sitter region and composite operators
of the infrared field theory.  We also point out that the truncation used
to find the reduced symmetry critical point can be extended to obtain a new
$\cN=4$ gauged supergravity theory holographically dual to a sector of
$\cN=2$ gauge theories based on quiver diagrams.

We consider more general kink geometries and construct a c-function that is
positive and monotonic if a weak energy condition holds in the bulk gravity
theory.  For even-dimensional boundaries, the c-function coincides with the
trace anomaly coefficients of the holographically related field theory in
limits where conformal invariance is recovered.

\vskip .3in


\Date{\sl {April, 1999}}

%
\parskip=4pt plus 15pt minus 1pt
\baselineskip=15pt plus 2pt minus 1pt
%
\newsec{Introduction}
\seclab\Introduction

The conjecture \refs{\JMalda} of the equivalence of string theory on
$AdS_5 \times S^5$ to ${\cal N}=4$ supersymmetric Yang-Mills theory in
$(3+1)$-dimensions derived some of its initial motivation and
plausibility
from the $SU(2,2|4)$ superconformal invariance of both sides of the
equivalence. Optimistically, one might expect that the strategy for
computing Green's functions \refs{\SGIKAP,\WitHolOne} is more general,
and applies to any situation where a quantum field theory (QFT) on the
boundary of a spacetime can be related to properties of a quantum
gravity theory in the bulk. To test this expectation it would be good
to have an example of a QFT--gravity pair where the QFT is
non-conformal but nevertheless well-understood at least in some aspects
which can be probed from the gravity side.  Furthermore, the bulk
description should be in terms of a geometry in which the supergravity
approximation to string theory is uniformly applicable.

The main purpose of this paper is to provide such an example. The QFT will
be the ${\cal N}=1$ supersymmetric gauge theory obtained from ${\cal N}=4$
super-Yang-Mills by adding a mass to one of the three chiral adjoint
superfields. The mass breaks conformal invariance and drives a
renormalization group flow.  Using the methods developed by Leigh and
Strassler \refs{\ls} one can argue that the theory recovers superconformal
invariance in the infrared, and that in fact there is a line of fixed
points in the infrared as well as the ultraviolet.  The theory can be made
strongly coupled in both limits by making the ultraviolet 't~Hooft coupling
large, and so a supergravity description at large $N$ should be possible.

We believe that the supergravity dual of this field theory flow is the
kink solution that is discussed in section~\Superkink, and which
interpolates between the AdS geometries at two critical points of
five-dimensional ${\cal N}=8$ gauged supergravity
\refs{\GRWplb,\GRW,\PPvN}. The critical point describing the UV end of
the flow is the expected maximally supersymmetric vacuum of the bulk
theory, which is dual to ${\cal N}=4$ SYM. The kink itself preserves
four real supercharges, as does the field theory described in
the previous paragraph. In the IR limit the kink approaches the 
critical point with ${\cal N}=2$ bulk supersymmetry that was
discovered in \refs{\KPW}.

The most direct evidence for the correspondence between the field 
theory flow and the supergravity kink is that the symmetries match. 
Along the whole RG flow there is $SU(2) \times U(1)$ 
global symmetry, and this symmetry is also present in the kink.  The
AdS geometries at the endpoints imply conformal symmetry in the UV 
and IR limits of the field theory, and there is $SU(2,2|4)$ symmetry at 
the UV end and $SU(2,2|1)$ symmetry at the IR end.  Furthermore, 
the limiting behavior of the kink near its UV endpoint involves scalars
with the right quantum numbers and dimensions to be dual to the 
mass term for one adjoint chiral superfield.
There is additional 
evidence from the match of trace anomaly coefficients computed at both
endpoints in field theory and through the AdS/CFT correspondence.
The authors of \refs{\klm} independently considered the field theory
RG flow and the matching of trace anomaly coefficients.

To probe this correspondence more deeply, we compute mass eigenvalues
of all fields of gauged supergravity at the non-trivial supersymmetric
critical point. We use this and the results of
\refs{\SGIKAP,\WitHolOne} to obtain the scaling dimensions of the
corresponding gauge theory operators. The details are presented in
section 6 and in an appendix.
Boson and fermion operators neatly combine
into multiplets that are representations of the superalgebra
$SU(2,2|1)$ of $\cN=1$ superconformal symmetry in four dimensions. We
exhibit gauge invariant combinations of the massless superfields of the
gauge theory
whose scaling dimensions and $SU(2) \times U(1)$ quantum numbers 
precisely match the five short multiplets observed in supergravity.  
There are three additional long multiplets which complete the picture.
This detailed field-operator map
constitutes perhaps the strong evidence that the supergravity critical
point is indeed the holographic dual of the mass-deformed ${\cal N}=4$ 
theory.

Non-conformal examples of the bulk-boundary holographic relation already
exist in the literature (see for instance
\refs{\imsy,\WitHolTwo,\abks,\ktZero,\minConfine, \SfetsosKehagias,\gDil}),
but from the point of view of testing the duality most of these are vexed
either by an incomplete understanding of the boundary theory, or by a bulk
geometry where supergravity is unreliable due to tachyons, or large
curvatures, or a strong dilaton---or some combination of these.

Our example is closer to the work of \refs{\kw,\samson,\lopez,\GPPZ,\JDFZ},
and we draw on various aspects of these papers in our analysis.  In
\refs{\kw} it was proposed that a blowup of an $S^5/{\bf Z}_2$ orbifold,
deformed to a coset manifold $SO(4)/SO(2)$, described the flow of an ${\cal
N}=2$ super-Yang-Mills theory to an ${\cal N}=1$ infrared fixed point
through the addition of a mass term.  The dual field to this mass term was
a twisted string state, so a direct analysis of the flow geometry beyond
the level of topology seemed difficult.  Similar examples were considered
in \refs{\samson,\lopez}, and in \refs{\samson} it was suggested that the
flows could be described in terms of tensor multiplets of an ${\cal N}=4$
gauged supergravity.  This was motivated in part by the work of
\refs{\GPPZ,\JDFZ}, where anti-de Sitter vacua of ${\cal N}=8$, $d=5$
gauged supergravity discovered in \refs{\GRW} were re-examined as candidate
descriptions of non-supersymmetric fixed points of RG flows.  The main
example considered in \refs{\GPPZ,\JDFZ} was a $SU(3)$-symmetric point.
Second order differential equations were found for a five-dimensional flow
geometry inteperpolating between the maximally supersymmetric point and
the $SU(3)$-symmetric one.  The field theory interpretation was that
adding a mass to one of the four gauginos resulted in a flow to an infrared
fixed point of RG.

A general problem of studying RG flows through the bulk-boundary
correspondence is that the validity of supergravity depends on having
strong coupling in the gauge theory, so that results cannot be compared
with perturbative treatments of the renormalization group flow.  For
instance, it is difficult to verify from a field theory point of view any
of the properties of the $SU(3)$-symmetric critical point studied in
\refs{\GPPZ,\JDFZ}.  What allows us to make non-trivial field theory
predictions about the infrared fixed point is the supersymmetry of the
fixed point and the ${\cal N}=1$ supersymmetry that must be preserved
throughout the flow.  These predictions are analogous to those which have
been checked in the conifold examples \refs{\kw,\gEin,\samson,\lopez}.
What is new in our paper as compared to these is that we are able to give
an explicit five-dimensional supergravity description of a supersymmetric
flow geometry.

Before restricting our attention to a particular example, we consider in
section~\RGflowss\ some general properties of the supergravity kinks that
could be used to describe renormalization group flows.  We are able to
identify a monotonic function along the kink which interpolates between the
anomaly coefficients at one end and at the other. Using nothing more than
Einstein's equations and a weak energy condition it is possible to show
that such a function can always be found in any kink geometry with the
Poincar\'e symmetries of the boundary theory in flat space. The argument is
dimension-independent. Thus, on rather general grounds, the holographic
bulk-boundary correspondence implies a c-theorem, at least in the regime
where classical gravity is a reliable guide to the bulk physics.  In the
four-dimensional case, an equivalent c-function was discussed independently
in \refs{\GPPZ}, and its monotonicity was checked using the equations of
motion for the supergravity geometries considered there.

In section~\Superkink\ we find the supergravity kink corresponding to the
supersymmetric renormalization group flow described above. We do this by
demanding that one eighth of supersymmetry is unbroken throughout the bulk,
and we thus obtain first order equations for a kink solution that
interpolates between the maximally supersymmetric critical point and the
one-quarter supersymmetric point of \refs{\KPW}. The first order
equations for the scalar fields are the gradient flow equations of a
``superpotential'' defined on a restricted four-dimensional slice of the
scalar manifold, and simply related to the potential of gauged
supergavity on this slice.

We work within the framework of gauged
$\cN=8$ supergravity in five dimensions because its full non-linear 
structure is known.  This is not the case for the complete Kaluza-Klein
reduction of ten-dimensional type~IIB supergravity on $S^5$.
The spectrum of five-dimensional ${\cal N}=8$ gauged supergravity includes 
only the first few Kaluza-Klein modes of the ten-dimensional theory, but
it is a complete theory at the classical level and was argued in 
\refs{\KPW} to be a 
consistent truncation of the parent theory.    
This has not been proven explicitly but is expected after suitable 
field redefinitions of the heavy KK modes.  Consistent truncation means 
that any solution of the truncated theory can be lifted to a solution of
the untruncated theory.

Another reason for working with five-dimensional gauged supergravity is
that it efficiently encodes a class of metric deformations of the
$S^5$, and a family of backgrounds for $B^{NS,RR}_{MN}$ fields where
the indices lie in the $S^5$ directions. To be more specific, the
five-dimensional gauged supergravity has 42 scalar fields that
parameterize the coset $E_{6(6)}/USp(8)$. From the holographic
perspective two of these scalars correspond to the Yang-Mills gauge
coupling and $\theta$-angle, 20 of them parametrize Yang-Mills scalar
masses (deformations of the metric on $S^5$), and the other 20
parametrize the possible Yang-Mills fermion masses (special backgrounds
for $B^{NS,RR}_{MN}$). Indeed gauged $\cN=8$ supergravity in five
dimensions is holographically dual to the Yang-Mills energy-momentum
tensor supermultiplet, and the supergravity scalars are
the moduli of this supermultiplet: the couplings and the masses. 

The lagrangian of gauged $\cN=8$ supergravity has a scalar potential
$V$ which is invariant under the gauged subalgebra, $SO(6)=SU(4)$, of
$E_{6(6)}$. Critical points of the potential $V$ give rise to $AdS_5$
solutions of gauged supergravity. The list of known critical points can
be found in \refs{\KPW}. For some of them the corresponding
ten-dimensional geometry can be found in the literature: for the
$SU(3)$ symmetric points see \refs{\Romans}; for the $SO(5)$ symmetric
points see \PvNNW. The critical point which we claim describes the
infrared endpoint of our flow is the recently discovered $SU(2) \times
U(1)$-symmetric point with one-quarter supersymmetry \refs{\KPW}. The
corresponding ten-dimensional geometry is, as yet, not known, but its
metric can be computed using the results of \KPW.

The known critical points of \KPW\ were found by specializing to an
eleven-dimensional, truncated submanifold, $\cS$, of $E_{6(6)}/USp(8)$.
This is the invariant subspace under a particular
$SU(2)$ subgroup of $SO(6)=SU(4)$.  Since this scalar submanifold
will play a major role in our work we will review its structure
in detail in section 3 and in two appendices.

In section~\NFourSUGRA\ we point out that the truncation of the scalar
manifold can be extended to the full theory and leads to a complete
five-dimensional $\cN=4$ gauged supergravity theory coupled to two tensor
multiplets.  We suggest that this theory captures a sector common to all
${\cal N}=2$ superconformal gauge theories based on quiver diagrams
\refs{\MDGM,\LNV}, and that the kink we have found may reflect certain
renormalization group flows in these theories as well.

A comment on the counting of supersymmetries may prevent future
confusion. A five-dimensional supergravity theory is referred to as having
${\cal N}$-extended supersymmetry if the invariance of the lagrangian is
characterized by $4{\cal N}$ real supercharges, and similarly with a
background of supergravity. The minimally supersymmetric five-dimensional
supergravity has ${\cal N}=2$ in the lagrangian. As always, maximal
supersymmetry is 32 real supercharges. A four-dimension field theory with
${\cal N}$-extended supersymmetry also has $4{\cal N}$ real supercharges,
unless it also has conformal symmetry, which doubles the number of
supercharges to $8{\cal N}$. In an effort to speak French to the French and
English to the Americans, we will
refer to a supergravity background dual to an ${\cal N}=1$ superconformal
field theory as ${\cal N}=2$.

\newsec{Field theory motivation}
\seclab\FieldTheory

Asymptotic freedom and confinement are difficult to capture reliably in
a supergravity geometry. For asymptotic freedom  this is
because the 't~Hooft coupling eventually gets weak in the ultraviolet,
and $\alpha'$ corrections become important to any supergravity
description. All geometric attempts to describe confinement include
singularities (some more malignant than others), and there is a variety
of problems regarding the discrepancy between the confining string
tension and the mass gap, nearly flat Regge trajectories, and unwanted
global symmetries.  The study of renormalization group
flows in supersymmetric gauge theories provides us with some examples
where both the ultraviolet and the infrared fixed points are conformal,
and neither asymptotic freedom nor confinement is encountered.

The example upon which we want to focus is a relevant deformation of
${\cal N}=4$ super-Yang-Mills theory by a mass term for one of the
three ${\cal N}=1$ adjoint chiral superfields which, together with the
${\cal N}=1$ adjoint vector superfield, fill out the full ${\cal N}=4$
gauge multiplet. In ${\cal N}=1$ language, the superpotential is
\eqn\SuperPot{
   W = \tr \Phi_3 [\Phi_1,\Phi_2] + \tf{1}{2} m \tr \Phi_3^2 \ .}
 The first term is the superpotential required by ${\cal N}=4$
supersymmetry. The second term breaks ${\cal N}=4$ down to ${\cal
N}=1$, and also destroys conformal invariance, so the total number of
real supercharges is reduced from thirty-two to four.  The theory does
have a non-trivial infrared fixed point.  Indeed, it was shown in \ls\
that by choosing the anomalous dimensions $\gamma_1 = \gamma_2 = -1/4$
and $\gamma_3 = 1/2$ for $\Phi_1$, $\Phi_2$, and $\Phi_3$, one could
simultaneously make the NSVZ exact beta function \refs{\NSVZ},
  \eqn\NSVZBeta{
   \beta(g) = -{g^3 N_c \over 8 \pi^2} {3 - \sum_i (1-2\gamma_i) \over
   1 - {g^2 N_c \over 8\pi^2}} \ ,}
 vanish, and make the superpotential in \SuperPot\ dimension three, which
is marginal. Note that it is immaterial in this analysis whether one
integrates out $\Phi_3$ or not.  The theory we have described has some
features in common with ${\cal N}=1$ SQCD with $N_f = 2N_c$, and it is its
own Seiberg dual.

That the infrared fixed point of this flow is the holographic dual theory
to the one-quarter supersymmetric vacuum found in \refs{\KPW} seems to have
been observed independently by several groups, including the present
authors, the authors of \refs{\klm}, and M.~Strassler
\StrasslerPrivate.  For us the crucial
observation was that the coefficients in the trace anomaly predicted by
field theory match those observed in supergravity. The coefficients we aim
to calculate are $a$ and $c$ in the one anomalous point functions which in
the notation of \afgj\ read
\eqn\TRAnomalies{\eqalign{
   \langle T^\mu_\mu \rangle_{g_{\mu\nu},V_\mu} &=
    {c \over 16 \pi^2} W_{\mu\nu\rho\sigma}^2 -
    {a \over 16 \pi^2} \tilde{R}_{\mu\nu\rho\sigma}
     \tilde{R}^{\mu\nu\rho\sigma} +
    {c \over 6 \pi^2} V_{\mu\nu}^2  \cr
   \langle \partial_\mu \sqrt{g} R^\mu \rangle_{g_{\mu\nu},V_\mu} &=
    -{a-c \over 24 \pi^2} R_{\mu\nu\rho\sigma}
     \tilde{R}^{\mu\nu\rho\sigma} +
    {5a-3c \over 9\pi^2} V_{\mu\nu} \tilde{V}^{\mu\nu}
  }}
in the presence of a metric $g_{\mu\nu}$ and a source $V_\mu$ for the
R-current $R^\mu$. The normalization of $a$ and $c$ is such that
$c=1/120$ for a single free, real scalar, and $c = (N_c^2-1)/4$ for
$SU(N_c)$ super-Yang-Mills. The combinations $a-c$ and $5a-3c$ in the
second line are consequences of ${\cal N}=1$ supersymmetry \afgj.
In the ultraviolet, where we can effectively set $m_3 = 0$, we arrive
at $a-c=0$ and $5a-3c \propto {8 \over 9} (N_c^2-1)$ by inspection of
the triangle diagrams in Figure~1. The current $R_\mu$ assigns
$r(\lambda) = 1$ to the gauginos and $r(\psi) = -1/3$ to the quarks.

\goodbreak\midinsert
\vskip .5cm
\centerline{ {\epsfxsize 4in\epsfbox{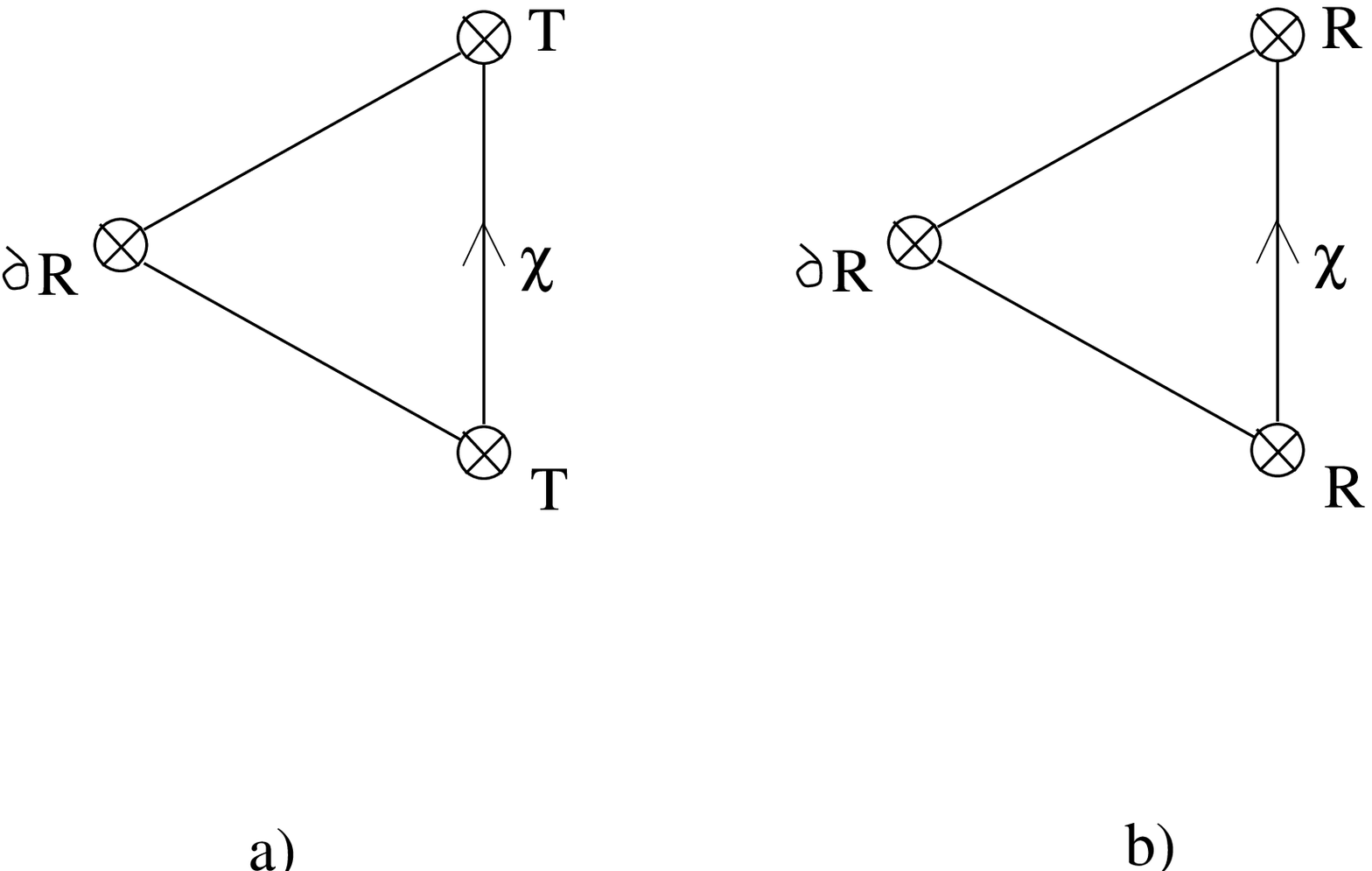}}}
   \vskip-2cm
   \centerline{$\displaystyle{a-c \propto \sum_\chi r(\chi)
     \qquad\qquad\qquad\quad
    5a-3c \propto \sum_\chi r(\chi)^3}$}
   \vskip1.5cm
\leftskip 2pc
\rightskip 2pc\noindent{\ninepoint\sl \baselineskip=8pt {\bf Fig.~1}:
Triangle diagrams for computing the anomalous contribution to
$\partial_\mu R^\mu$.  The sum is over the chiral fermions $\chi$ which
run around the loop, and $r(\chi)$ is the R-charge of each such
fermion.
}
\endinsert
When $m \neq 0$, this R-current becomes anomalous, which is entirely
appropriate since $\partial_\mu R^\mu$ is an ${\cal N}=1$ superpartner
of $T^\mu_\mu$, and the theory now is changing with scale. There is
however a non-anomalous combination \refs{\ksv} of $R_\mu$ with the Konishi
currents, $K^i_\mu$.  By definition, $K^i_\mu$ assigns charge~$1$ to the
fields in the $i^{\rm th}$ chiral multiplet and charge~$0$ to the
fields in vector multiplets.  The non-anomalous current is
  \eqn\SDef{
   S_\mu = R_\mu + \tf{2}{3} \sum_i
    \left( \gamma_{\rm IR}^i - \gamma^i \right) K^i_\mu \ .}
 Here $\gamma^i$ is the anomalous dimension of the $i^{\rm th}$ chiral
superfield, and the values of $\gamma^i_{\rm IR}$ are those discussed in
the text preceding \NSVZBeta. The 't~Hooft anomaly matching implies that
$\langle \partial_\mu R^\mu \rangle$ in the infrared is equal to $\langle
\partial_\mu S^\mu \rangle$ in the ultraviolet.  In the ultraviolet,
$\gamma^i = 0$ and the perturbative analysis in terms of fermions running
around a loop can be applied. The result is $a-c=0$ and $5a-3c = {3 \over
4} (N_c^2-1)$ with the same constant of proportionality as before.
Combining with the results of the previous paragraph, one finds
\eqn\ChargeRatio{
   {a_{\rm IR} \over a_{\rm UV}} = {c_{\rm IR} \over c_{\rm UV}}
     = {27 \over 32} \ .}
The non-anomalous $S_\mu$ current generates an exact symmetry of the
mass-deformed theory, and it will be matched precisely by a $U(1)$
symmetry of the supergravity kink solution. This symmetry is called
$U(1)_R$ in subsequent sections.

{}From the gravity side it is more straightforward to compute $\langle
T^\mu_\mu \rangle$, since we can rely on the analysis of \MHKS.  We remain
in mostly minus conventions.  Suppose we are given any compactification of
string theory or M-theory (or any other, as-yet-unknown theory of quantum
gravity) whose non-compact portion is $AdS_5$ with $R_{\mu\nu} = \Lambda
g_{\mu\nu}$.  If we rescale the metric by a factor of $4/\Lambda$, we
obtain the dimensionless $AdS_5$ metric $\widehat{ds}_5^2$ with
$\hat{R}_{\mu\nu} = 4 \hat{g}_{\mu\nu}$. In defining a conformal field
theory through its duality to the $AdS_5$ compactification under
consideration, the part of the action relevant to the computation of
$\langle T^\mu_\mu \rangle$ is the Einstein-Hilbert term plus the
cosmological term:
\eqn\CosmConstC{
   S = -{1 \over 2\kappa_5^2} \int d^5 x \, \sqrt{g}
    \left( R - 3\Lambda + \ldots \right)
     = -{4 \over \kappa_5^2 \Lambda^{3/2}} \int d^5 x \, \sqrt{\hat{g}}
    \left( \hat{R} - 12 + \ldots \right) \ ,}
where $\kappa_5^2 = 8\pi G_5$ is the five-dimensional gravitational
coupling.  Directly from \MHKS\ we read off
\eqn\acCos{ a = c = {\pi \over G_5 \Lambda^{3/2}} \ . }
 In solutions of ${\cal N}=8$ supergravity where the scalars are fixed,
$\Lambda = -{4 \over 3} V$.  It was commented on in \refs{\KPW} already
that
  \eqn\acRatioAgain{
   {c_{\rm IR} \over c_{\rm UV}} =
    \left( {V_{\rm UV} \over V_{\rm IR}} \right)^{-3/2} =
    {27 \over 32} \ .}
 The agreement betwen \ChargeRatio\ and \acRatioAgain\ is a sign that we
have found the correct field theory interpretation of the infrared AdS
region.  It is straightforward to check the overall normalization of $a$
and $c$: \acCos\ leads to $a_{UV} = c_{UV} = N^2/4$, which differs from the
$SU(N)$ gauge theory result only by a $1/N^2$ correction.  This agreement
can be traced back to absorption calculations \refs{\gkSchwing}.

The essentials of the anomaly-matching analysis were worked out in
\gEin\ for the conifold theory introduced in
\refs{\kw}. Aspects of it were also employed in \refs{\JDFZ} and in
particular in \refs{\klm} in a field theory analysis independent of our
own for the one-quarter supersymmetric fixed point of \refs{\KPW}.

The fields involved in the interpolating solution between the maximally
supersymmetric anti-de Sitter vacuum of gauged supergravity and the
one-quarter supersymmetric vacuum have masses near the maximally
supersymmetric end which correspond to operators of dimensions $2$ and
$3$. These are, respectively, the boson and fermion mass involved in the
second term of \SuperPot.  Appropriately, turning on the scalar dual to the
boson mass term preserves $SO(4) \times SO(2) \subset SO(6)$, while turning
on the scalar dual to the fermion mass term preserves $SU(3)$.  Together
they break the symmetry down to $SU(2) \times U(1)$.  If one regards the
presence of these scalars in the interpolating solution, plus the
supersymmetry, plus the existence of the interior AdS region, as sufficient
evidence that the infrared fixed point in field theory is described by the
deformation of ${\cal N}=4$ super-Yang-Mills which we have discussed above,
then the agreement between \ChargeRatio\ and \acRatioAgain\ can be regarded
as a successful field theory prediction of the value of the scalar
potential $V$ at the saddle point that defines this solution. From this
point of view, testing dimensions of scalar operators in field theory
against masses of gauged supergravity fields amounts to checking field
theory predictions of the second derivatives of $V$ at the saddle point.

\newsec{$\cN=8$ Supergravity and conformal phases of Yang-Mills theory}
\seclab\SugraandYM

All the $42$ scalars in five-dimensional ${\cal N}=8$ gauged supergravity
participate in the scalar potential except for the dilaton and the axion.
In view of the correspondence to operators in the Yang-Mills theory
explained in the introduction, critical points of this potential correspond
to conformal fixed points of $\cN =4$ Yang-Mills theory obtained by turning
(Yang-Mills) fermion and scalar masses and running to the infrared.

In an effort to classify some of the new phases of the Yang-Mills
theory the authors of \KPW\ classified all critical points of the
$\cN=8$ gauged supergravity potential preserving at least a particular
$SU(2)$ symmetry.  The initial motivation for this was largely
pragmatic: reducing the problem of forty two scalars to a solvable one
involving only eleven.  As we will explain in more detail in
section~\NFourSUGRA, the truncation to $SU(2)$ singlets can be
extended to the whole ${\cal N}=8$ theory, and the result is a gauged
${\cal N}=4$ supergravity theory coupled to two tensor multiplets.  We
believe that this truncated theory is holographically dual to a
subsector of operators which is common to all ${\cal N}=2$
super-Yang-Mills theories based upon quiver diagrams.

The purpose of this section is to review the analysis in \KPW\ of
critical points in supergravity with particular attention to the one
that preserves one quarter of supersymmetry.  This critical point is
a reasonable candidate for the description of the IR fixed point of
the field theory flow described in section~\FieldTheory.  We begin in
section~\ScalarStructure\ by summarizing the $SU(2)$ singlet sector of
the supergravity scalars.  We also define a sort of superpotential
which is helpful for the analysis of supersymmetry.  We summarize this
analysis at the critical point in section~\SuperBackgrounds, and we
extend it to non-anti-de Sitter geometries in section~\Superkink.

\subsec{The scalar structure of the $SU(2)$ singlet sector}
\subseclab\ScalarStructure

The $\cN=8$ supergravity theory has a gauge symmetry of $SU(4)$ or
$SO(6)$.  We consider the $SU(2)_I \times SU(2)_G \times U(1)_G$
subgroup of $SU(4)$, and look at the  singlets of $SU(2)_I$.
(The subscripts $I$ and  $G$ on the groups are intended to
distinguish the ``Invariance''  group from the residual
``Gauge'' group.)   As was discussed in \KPW, these
scalar singlets may be thought of as parametrizing the coset:
\eqn\Scoset{\cS ~=~ {SO(5,2) \over SO(5) \times SO(2)} ~\times~
O(1,1)\ .}
The group $SO(5,2)$ contains an obvious $SO(3) \times SO(2,2)$
subgroup, and $SO(2,2) = SL(2,\IR) \times SL(2,\IR)$.
The $SO(3)$ is $SU(2)_G$, one $SL(2,\IR)$ describes the dilaton
and axion of the theory, and $U(1)_G$ is the compact subgroup
of the other $SL(2,\IR)$.

The scalar potential, $V$, on the eleven-dimenensional scalar manifold,
${\cal S}$, has a residual $SU(2) \times U(1) \times SL(2,\IR)$
invariance and can therefore be parametrized in terms of four real
variables \refs{\KPW}:
\eqn\potredcd{\eqalign{
V ~=~ -{g^2 \over 4}~\Big[  & \rho^{-4}~ \big( 1 - \cos^2(2 \phi)~
(\sinh^2(\varphi_1) - \sinh^2(\varphi_2))^2 \big) \cr & ~+~
\rho^2(\cosh(2 \varphi_1) + \cosh(2 \varphi_2) ) ~+   {1 \over 16}
\rho^8 ~\big(2 + 2 \sin^2(2 \phi) \cr & ~-~ 2 \sin^2(2 \phi)
\cosh(2 (\varphi_1- \varphi_2)) - \cosh(4 \varphi_1) - \cosh(4
\varphi_2) \big)\Big] \ .}}
 Our variables are related to those of \KPW\ by $r_x = -(\varphi_1-
\varphi_2)/2$, $r_y = (\varphi_1+\varphi_2)/2$, and $\theta = 2 \phi$.  As
in \KPW, $\rho = e^\alpha$, parametrizes one sheet of the $O(1,1)$ factor
of \Scoset.

The $\cN=8$ supergravity has a $USp(8)$ invariance, and in particular
the scalar structure is encoded in two  $USp(8)$ tensors, $W_{ab}$ and
$A_{abcd}$.  The truncation of the $USp(8)$ indices to the
singlets of $SU(2)_I$ can be effected by projecting the ${\bf 8}$
of $USp(8)$ onto the space spanned by the four vectors
\eqn\USPevecs{\eqalign{v_1 ~=~ & (1,0,0,0,0,1,0,0)\ , \qquad
\tilde v_1 ~=~  (0,1,0,0,-1,0,0,0) \ , \cr v_2 ~=~ & (0,0,0,1,0,0,1,0)
\ , \qquad  \tilde v_2 ~=~  (0,0,1,0,0,0,0,-1) \ .}}
These vectors also turn out to be global eigenvectors of the
tensor $W_{ab}$, with eigenvalues, $\lambda_1, \bar \lambda_1$
and  $\lambda_2, \bar \lambda_2$ respectively, where
\eqn\Wevals{\eqalign{\lambda_1 ~=~ -{e^{-2 i \phi} \over 4 \rho^2}~
\Big[ &\rho^{6}~ \big( 2 \cos(2 \phi)~-~ \cosh(2 \varphi_1) ~+~
\cosh(2 \varphi_2) ~+~ 2 i \sin(2 \phi) \cosh (\varphi_1- \varphi_2)
\big) \cr & ~+~ \big( 2 \cos(2 \phi) (\cosh(2 \varphi_1) +
\cosh(2 \varphi_2)) ~+~ 4 i \sin(2 \phi) \cosh(\varphi_1+ \varphi_2)
\big) \Big] \ , \cr \lambda_2 ~=~ -{e^{-2 i \phi}
\over 4 \rho^2}~  \Big[ &\rho^{6}~  \big( 2 \cos(2 \phi)~+~
\cosh(2 \varphi_1) ~-~  \cosh(2 \varphi_2) ~+~ 2 i \sin(2 \phi) \cosh
(\varphi_1- \varphi_2) \big) \cr & ~+~  \big( 2 \cos(2 \phi)
(\cosh(2 \varphi_1) + \cosh(2 \varphi_2)) ~+~
4 i \sin(2 \phi)  \cosh(\varphi_1+ \varphi_2) \big) \Big]  \ .}}

These eigenvalues also provide a superpotential $W$ related to $V$ by
\eqn\VfromW{V ~=~ {g^2 \over 8}~\sum_{j = 1}^3 ~\Big| {\del W
\over \del \varphi_j} \Big|^2 ~-~ {g^2 \over 3}~\big|W \big|^2 \ ,}
where $\varphi_3 = \sqrt{6}~\alpha$, and $W = \lambda_1$ or $W =
\lambda_2$.  We use the metric $\delta_{ij}$ to contract the
indices of the partial derivatives of $W$ with respect to $\varphi_i$, 
and as we will see,
this is the proper metric on this part of the scalar manifold.

The scalar and gravity part of the $\cN = 8$ supergravity action
is \GRW:
\eqn\SGAction{\int d^5 x~ \sqrt{|g|}~
\Big[ - \coeff{1}{4} ~R ~+~ \coeff{1}{24}~g^{\alpha \beta} ~
P_{\alpha\, abcd} P_\beta{}^{abcd} {}~-~ V \Big]\ .}
In \SGAction\ and in the rest of this paper, we are working in
five-dimensional Planck units such that $\kappa_5^2 = 2$.  The efficient
way to repristinate Newton's constant is to insert $2/\kappa_5^2$ as 
an overall factor in from of \SGAction: that way the equations of motion 
are not changed. In particular, the maximally supersymmetric $AdS_5$
vacuum of the theory is determined by the equation
\eqn\NFourPoint{
 R_{\alpha\beta} = -{4 \over 3}~V_0~g_{\alpha\beta}
= g^2 g_{\alpha\beta} \ .}
Comparison with the usual form, $R_{\alpha\beta} = {4 \over L^2}
g_{\alpha\beta}$, yields $g = 2/L$. In the $AdS_5 \times S^5$ vacuum
of type~IIB theory supported by $N$ units of self-dual five-form flux, $L$
enters as the radius of the $S^5$, and one has the standard relation
\eqn\LKappa{L^4 = {\kappa_{10} N \over 2\pi^{5/2}} \ .}
In the foregoing parametrization of the scalar manifold, the
scalar kinetic term of \GRW\ reduces to:
\eqn\scalarkin{\eqalign{{1 \over 24}~g^{\mu \nu}~\big(&
P_{\mu\, abcd}~ P_\nu{}^{abcd}\big) ~=~  \cr & {1 \over 2}~
\Big[~ \sum_{j = 1}^3 ~ g^{\mu \nu}~ (\del_\mu \varphi_j) ~(\del_\nu
\varphi_j) ~\Big] ~+~  \sinh^2(\varphi_1 - \varphi_2)~g^{\mu \nu}~
(\del_\mu \phi) ~ (\del_\nu \phi) \ .}}
which indeed shows that the metric on the scalar
fields $\varphi_j$ is simply $\delta_{ij}$.

\subsec{Supersymmetric backgrounds}
\subseclab\SuperBackgrounds

To find supersymmetric bosonic backgrounds one sets the
variations of the spin-$1/2$ and spin-$3/2$ fields to zero.
{}From \GRW, the gravitational and scalar parts of these
variations are:
\eqn\fermvars{\eqalign{\delta\psi_{\mu a} ~=~ &{\cal D}_\mu
\epsilon_a ~-~ \coeff{1}{6} g W_{ab} \gamma_\mu \epsilon^b \ ,
\cr \delta\chi_{abc} ~=~ &\sqrt{2}~\Big[\gamma^\mu
P_{\mu\, abcd} \epsilon^d ~-~ \coeff{1}{2} g A_{dabc}
\epsilon^d \Big]\ .}}
 We remind the reader that we use mostly minus metric conventions,
and $\{\gamma^\mu,\gamma^\nu\} = 2 \eta^{\mu\nu}$.
For a background with constant scalars, one finds that the vanishing of
$\delta\psi$ relates the AdS radius to the eigenvalues of $W_{ab}$. The
vanishing of $\delta\chi_{abc}$ is equivalent, via a tensor identity,
to the vanishing of $\delta \psi_{\mu a}$.  However, vanishing of
$\delta\chi_{abc}$ directly leads to the condition that one has a
supersymmetric background with constant scalars if and only if $\phi
=0$ and $\del W/\del \varphi_j =0$ in \VfromW. This statement was
proved
with the help of Mathematica calculations, and is the first of several
important places in this work where Mathematica was essential.
Note that the conditions $\phi=0$ and $\del W/\del \varphi_j =0$ are
sufficient but not necessary condition for $V$ to have a critical
point.

One can easily verify that supersymmetry preserving
ground states are obtained by taking:
\eqn\susyGS{\varphi_1 ~=~ \pm \coeff{1}{2}~\log(3) \ , \quad
 \varphi_2 ~=~ 0, \quad \varphi_3 \equiv \sqrt{6}~ \alpha ~=~
\coeff{1}{\sqrt{6}} \log(2) \ , \quad \phi = 0 \ .}
 All other $SU(2)_I$-invariant supersymmetric ground states (besides the
trivial ${\cal N}=4$ point) are $\ZZ_2$ images of the ones specified in
\susyGS, and can be obtained from these either via the interchange
$\varphi_1 \leftrightarrow \varphi_2$ or by going to the other component of
$O(1,1)$.
These supersymmetric ground states were found in
\KPW.  They all have $V_0 = -{2^{4/3} \over 3} g^2$
and $\cN=2$ supersymmetry (in the supergravity sense described at the end
of section~\Introduction).  The supersymmetry generators for \susyGS\ are
given by using the projectors $v_1$ and $\tilde v_1$ of \USPevecs.  If one
considers the ground states with $\varphi_1 \leftrightarrow \varphi_2$,
then the unbroken supersymmetries are given by $v_2$ and $\tilde v_2$.

In section~\sugraandopmap\ we will investigate the supergravity
spectrum at the critical point specified by \susyGS. Our results, and
their interpretation within the holographic map between bulk field and
boundary operators, will provide detailed evidence that this critical
point is the supergravity description of the infrared fixed point of
the supersymmetric RG flow described in section~\FieldTheory. The {\it
prima facie} evidence for this, as commented in section~\FieldTheory,
is that the value of $V_0$ translates into the correct trace anomaly
for the supersymmetric gauge theory. Before entering into the detailed
investigation of the fixed point, we will show how supergravity can
provide a description of the entire RG flow from the ${\cal N}=4$ UV
theory to the ${\cal N}=1$ infrared fixed point.

\newsec{Generalities on holographic RG flows: a c-theorem}
\seclab\RGflowss

An important aspect of the AdS/CFT correspondence is the notion that the
radial coordinate $U$ of AdS can be regarded as a measure of energy.  Thus
a geometrical cutoff of AdS serves as an ultraviolet regulator.  These
ideas were implicit in \refs{\JMalda,\SGIKAP,\WitHolOne} and have
subsequently been developed more fully in \refs{\WitSuss,\PeetPolch}.
Phenomena that probe smaller values of $U$ are to be thought of as probing
the infrared structure of the theory. For instance, if one computes Green's
functions by setting the values of supergravity fields at finite $U$, then
making $U$ smaller is believed to provide a natural coarse-graining of the
UV details, thereby implementing a version of the Wilsonian renormalization
group.

In this context a supergravity ``kink'' in the $U$-direction
interpolating between $U=\infty$ and $U=0$ can be thought of as an
explicit construction of a renormalization group flow between a UV
fixed point and an IR fixed point of the ``boundary'' field theory
\refs{\GPPZ,\JDFZ}. Our purpose in this section is to exhibit the
general ansatz for a kink and to demonstrate one simple property of
kinks that translates via holography into a c-theorem for the boundary
theory.

\subsec{The kink ansatz}
\subseclab\KinkAnsatz

We start by considering arbitrary bulk dimension $D$.  We are
interested in boundary theories that are Poincar\'e invariant in
$(D-1)$~dimensions.  The most general possible $D$-dimensional bulk
metric consistent with this symmetry can be written as
\eqn\RGFmetric{
ds^2 = e^{2 A(r)} \eta_{\mu\nu} dx^\mu dx^\nu - dr^2 \ .}
We will always use the mostly minus metric convention, which has the
consequence that anti-de Sitter space has a positive cosmological
constant. We could generalize \RGFmetric\ by replacing $\eta_{\mu\nu}$
by an arbitrary Ricci-flat metric $\gamma_{\mu\nu}$, and none of the
analysis in this section or the next would change. If $A(r) = {r \over
\ell}$, then \RGFmetric\ becomes anti-de Sitter space with
$R_{\alpha\beta} = {D-1 \over \ell^2} g_{\alpha\beta}$. Other standard
radial variables for anti-de Sitter space are $U = e^{r/\ell}/\ell$
\refs{\JMalda} and $z = x_0 = \ell e^{-r/\ell}$
\refs{\SGIKAP, \WitHolOne}. The kink that we shall exhibit in
section~\Superkink\ interpolates between an $r \to \infty$
``ultraviolet'' $AdS_5$ region, where $A(r)$ is linear, and an $r \to
-\infty$ ``infrared'' region, where $A(r)$ is again linear, but with a
larger positive slope. A sketch of the spacetime with the spatial boundary 
dimensions suppressed is shown in Figure~2.

The geometry depicted in Figure~2 is not the only possibility allowed by
the ansatz \RGFmetric. For the purpose of studying the AdS/CFT
correspondence we are interested in geometries which are asymptotic to
anti-de Sitter space on one end: that is, $A(r) \sim {r \over \ell}$ as $r
\to \infty$.  Other geometries are possible where curvature singularities
appear at finite $r$ \refs{\SfetsosKehagias,\gDil,\PorratiConfine}.
It is also possible that the geometry will have no singularities that
are causally connected to the boundary, but that $A(r)$ never recovers
linear behavior. The one general restriction, which we will prove in
section~\CTheorem, is $A''(r) \leq 0$. This rules out a second anti-de
Sitter boundary: it is impossible to have $A'(r)$ approach a positive
constant as $r \to \infty$ and a negative constant as $r \to
-\infty$. The monotonicity of $A'(r)$ implies a c-theorem, as we shall
see.

\goodbreak\midinsert
\vskip .5cm
\centerline{ {\epsfxsize 4in\epsfbox{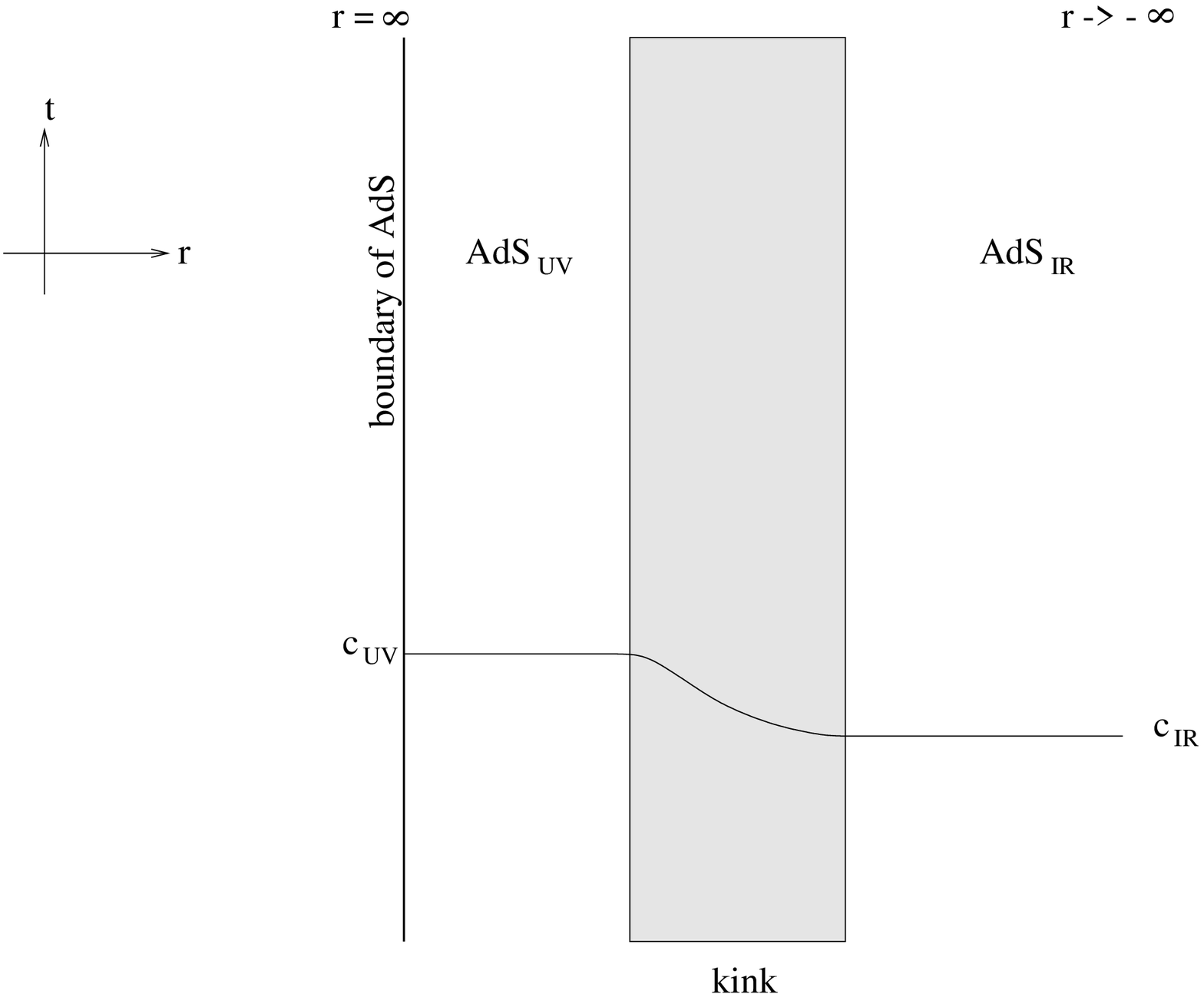}}}
\leftskip 2pc
\rightskip 2pc\noindent{\ninepoint\sl \baselineskip=8pt
{\bf Fig.~2}:
The supergravity kink interpolates
between an asymptotically AdS space in the UV ($r \to
\infty$) and another in the IR ($r \to - \infty$).  The value of
${\cal C}(r)$ decreases monotonically as one moves toward the
infrared, and in the asymptotic regions the value of
${\cal C}(r)$ approaches the central charge.
}
\endinsert

\subsec{A c-theorem}
\subseclab\CTheorem

The analysis of anomaly coefficients in \refs{\MHKS} applies to anti-de
Sitter spaces $AdS_D$ for any odd bulk dimension $D$. The anomalous
value of $\langle T^\mu_\mu \rangle$ appears in the supergravity
analysis because the cutoff $r = r_0$ does not respect diffeomorphism
invariance. Different choices of radial variable (including those which
mix the coordinates $r$ and $x_\mu$ in \RGFmetric)
correspond to different but conformally equivalent boundary metrics
$\gamma_{\mu\nu}$. The general result of \refs{\MHKS} is
  \eqn\AnyAnom{
   \langle T^\mu_\mu \rangle = {\hbox{universal} \over A'^{D-2}} \ ,
  }
where the universal part is the same for all theories, and is a curvature
invariant involving $D-1$ derivatives of the boundary metric.  No such
invariant exists in odd boundary dimensions.  The denominator of \AnyAnom\
actually appears in the analysis of \refs{\MHKS} as the inverse radius of
the anti-de Sitter space, and no particular coordinate choice is
implied. Expressing this inverse radius as $A'$ is merely a convenience.

The metric
\eqn\RGFmetricAgain{ds^2 ~=~ e^{2 A(r)} ~ (~\gamma_{\mu\nu} d x^\mu
d x^\nu~) ~-~ dr^2 \ ,}
 where $\gamma_{\mu\nu}$ is Ricci flat, has a Ricci tensor whose
nonvanishing components are
  \eqn\Ricci{R_{\mu \nu} ~=~ e^{2 A(r)}~\big[ A'' ~+~ (D-1)
  (A')^2 \big] ~\gamma_{\mu\nu} \ , \qquad R_{r r} ~=~-(D-1)
  \big[ A'' ~+~ (A')^2 \big] \ .}
 We note that
  \eqn\Punchline{
   -(D-2) A'' = R^t_t - R^r_r = G^t_t - G^r_r =
     \kappa_D^2 (T^t_t - T^r_r) \ ,
  }
 where $\kappa_D$ is the $D$-dimensional gravitational constant and in the
last equality we have used Einstein's equations, $G_{\alpha\beta} =
\kappa_D^2 T_{\alpha\beta}$.  Note that we are including any cosmological
constant as a term in $T_{\alpha\beta}$ proportional to $g_{\alpha\beta}$.
We would now like to argue that the last expression in \Punchline\ has to
be nonnegative.  Without resorting to a particular lagrangian, we can try
to appeal to some appropriate positive energy condition.

The stress tensor that supports the geometry \RGFmetricAgain\ is diagonal:
  \eqn\DiagFormT{
   T^\alpha_\beta = \diag\{ \rho,-p_1,-p_2,\ldots,-p_{D-2},-p_{D-1} \} \ ,
  }
 where $p_{D-1} = p_r$.  There are four different energy conditions in
common use: for $D>2$, they read \refs{\Wald}\foot{The form we quote for
the strong energy condition is the natural one given its motivation in
terms of the ricci tensor: $R_{\alpha\beta} \xi^\alpha \xi^\beta \geq 0$
for timelike or null vectors $\xi^\alpha$.}
  \eqn\EnergyConditions{
   \seqalign{\span\TT\quad & \span\TR \quad\quad & \span\TR}{
    Energy condition & \hbox{General form} & \hbox{Diagonal 
     $T^\alpha_\beta$}  \cr
    Strong: & \textstyle{
     \!\left( T_{\alpha\beta} - {1 \over D-2} g_{\alpha\beta} 
     T \right) \xi^\mu \xi^\nu \geq 0} & 
     \rho + p_i \geq 0,  \cr\noalign{\vskip-1\jot} & & 
     \textstyle{\quad (D-3)\rho + \sum_i p_i \geq 0}  \cr\noalign{\vskip-0.5\jot}
    Dominant: & T_{\alpha\beta} \xi^\alpha \eta^\beta \geq 0 & 
     \rho \geq |p_i|  \cr
    Weak: & T_{\alpha\beta} \xi^\alpha \xi^\beta \geq 0 & 
     \rho + p_i \geq 0,\ \ \rho \geq 0  \cr
    Weaker: & T_{\alpha\beta} \zeta^\alpha \zeta^\beta \geq 0 &
     \rho + p_i \geq 0,
  }}
 where $\xi^\alpha$ and $\eta^\alpha$ are arbitrary future-directed
timelike or null vectors and $\zeta^\alpha$ is an arbitrary null vector.
In the second column of \EnergyConditions\ we have given the general
definition of the energy condition, and in the third column we have
indicated how it constrains a diagonal stress-energy tensor.  What we have
called the weaker energy condition is implied by all the others, and what
we have called the weak energy condition is implied by the dominant energy
condition.  Otherwise there are no logical relations among the various
conditions.  With our convention that any cosmological constant should be
included in the stress tensor as a term proportional to $g_{\alpha\beta}$,
none of the energy conditions is satisfied by anti-de Sitter space, except
the weaker energy condition.

In our case, Poincar\'e invariance dictates $\rho + p_i = 0$ for $i =
1,2,\ldots, D-2$.  We would have $\rho + p_r = 0$ as well if and only if
the cosmological constant is the only contribution to the
stress-tensor---and then we recover perfect anti-de Sitter space, so there
is no RG flow at all.  What we learn, then, is that given Poincar\'e
invariance, the desired monotonicity relation $A'' \leq 0$ is exactly
equivalent to the weaker energy condition.\foot{In the initial version of
this paper we mistakenly used the dominant energy condition in place of the
weaker energy condition.  We thank M.~Porrati for private communications
regarding the relevance of the weaker energy condition which cleared up
this error.}

Consider an example: suppose we had an action
\eqn\genericSGA{{\cal I}_{SG} ~=~ {1 \over 2 \kappa_D^2}
 \int d^{D} x~ \sqrt{|g|}~
\Big[ - R ~+~ \half~g^{\alpha \beta} {\cal M}_{IJ} P_\alpha^I
P_\beta^J {}~-~ 4 V \Big]\ ,}
where $P_\alpha^I$ represents derivatives of {\it any} set of scalar
fields, ${\cal M}_{IJ}$ is some positive definite metric on the scalar
manifold, and $V$ is any scalar potential.  If we insist on Poincar\'e
invariance, then the scalars can only depend on $r$.  Then we find
  \eqn\ScalarCase{
   T^t_t - T^r_r = {1 \over 2 \kappa_D^2} {\cal M}_{IJ} P^I_r P^J_r \ ,
  }
which is positive because ${\cal M}_{IJ}$ is positive definite.  The
scalar potential's contribution to $T^t_t$ and $T^r_r$ cancels.

Our example is general enough to cover any flow in gauged supergravity
involving only the scalar fields and the metric. (It seems difficult to
preserve Poincar\'e invariance on the boundary if one turns on any bulk
matter fields with spin). But \Punchline, together with the assumption
of the weaker energy condition, constitutes an even more general
proof that the quantity
\eqn\CFunctionGen{
{\cal C}(r) = {{\cal C}_0 \over A'^{D-2}} }
is non-increasing along the flow toward the infrared. 
The parameter, ${\cal C}_0$, is a constant.

The result \AnyAnom\ shows that if the geometry is anti-de Sitter, then
the anomaly coefficients of the corresponding conformal field theory
are proportional to ${\cal C}$. With appropriate definitions of those
anomaly coefficients and of ${\cal C}_0$, we can say that ${\cal C}$
coincides with the anomaly coefficients for anti-de Sitter geometries.
In the duality of ${\cal N}=4$ $SU(N)$ super-Yang-Mills theory with
${\cal N}=8$ gauged supergravity, one can show that
\eqn\CFunctionKink{
   {\cal C}(r) = {g^3 N^2 / 32 \over A'(r)^3} }
is the appropriate normalization to give ${\cal C}=a=c={N^2 \over 4}$
for the unperturbed maximally supersymmetric anti-de Sitter vacuum of
the supergravity theory, and ${\cal C}=a=c={27 \over 128} N^2$ for the
one-quarter supersymmetric vacuum found in \refs{\KPW}. In
\CFunctionKink, $g$ is the gauge coupling in the five-dimensional
supergravity, and it has units of energy. The important thing to
realize is that once we have fixed ${\cal C}_0$ within a particular
supergravity theory, the formula \CFunctionGen\ gives a uniform
prescription for computing the anomaly coefficients in any conformal
theory dual to an anti-de Sitter vacuum of that supergravity theory.

Now we want to consider a kink spacetime which is asymptotic to anti-de
Sitter spaces for $r \to \pm\infty$. The analysis of \refs{\MHKS} does
not apply directly to such a spacetime. But its asymptotic limits as $r
\to \pm\infty$ are anti-de Sitter vacua, and we can think of obtaining
the UV and IR anomaly coefficients for these vacua from \CFunctionGen.
To apply the analysis of \refs{\MHKS} directly we must ``lift out'' one
asymptotically anti-de Sitter end from the kink and make it into an
exactly anti-de Sitter spacetime. But the answer for ${\cal C}$ is
exactly the same as if we had evaluated $\lim_{r \to \pm \infty} {\cal
C}(r)$ in the original non-anti-de Sitter kink geometry. This is
precisely analogous to the situation in field theory: properties
pertaining narrowly to a given fixed point do not depend on a
particular flow that leads into it or out of it; such properties can be
studied just as well by considering the trivial flow which remains
forever at the fixed point.

It is difficult or impossible to give a scheme-independent definition of
the central charge away from conformal fixed points.  Flows may exist which
pass arbitrarily close to one fixed point on their way to another.  For
these a ``scaling region'' exists where $A'$ is much larger than higher
derivatives of $A$.  The function, ${\cal C}(r)$, in such a region is 
close to the anomaly coefficient of the nearby fixed point.

In \Zamo, the monotonicity of the c-function followed trivially from
the fact that the RG flow was the gradient flow of the c-function. In
general it seems difficult to prove any analog of that statement in
four dimensions. However, for supersymmetric supergravity kinks, an
analogous statement can be proven: as we will explain in
section~\Superkink, $A'$ can be expressed as a function of the scalar
fields (without derivatives), and the trajectory of the scalars
throughout the flow is specified by gradient flow of that function.

Our proof of a holographic c-theorem also applies to flows such as the
one envisaged in \refs{\kw}, where light string states from twisted
sectors are involved. In addition, it would apply to flows involving 
any or all of the massive Kaluza-Klein states arising from any
compactification of ten-dimensional supergravity down to
five-dimensions. It even applies if the matter lagrangian includes
higher derivative terms, provided the weaker energy condition
continues to hold. However, the proof does rely on Einstein's
equations. It would be interesting to explore whether the argument can
be extended to cover the alterations in the gravitational action that
arise through $\alpha'$ corrections. We should also add that the
inequality $A'' \leq 0$ turned out to be {\it exactly} the weaker
energy condition on the stress-tensor supporting the geometry, so if
consistent matter violating the weaker energy condition exists in the
bulk in a regime where the classical gravity approximation is valid, it
seems likely that one could use it to construct holographically a
boundary theory which violates the c-theorem.

In summary, we
seem to have a dimension-independent proof of a c-theorem from the
weaker energy condition. It applies to theories where a gravity dual
can be found in which Einstein's equations hold, and in which the
gravity approximation applies. Besides the obvious large $N$, strong
coupling restrictions, this also amounts in four dimensions to the
restriction $a-c=0$, at least to leading order in large $N$. (The
$1/N$ corrections were explored in \refs{\AnselmiKehagias,\apty}.)
The proof of the theorem, equation \Punchline, is appropriately trivial
for such a general truth.

There have been other recent efforts toward a c-theorem in higher
dimensions: \refs{\ForteLatorre,\AnsAPrime} take a field theoretic
approach, \refs{\Appelquist} works with thermodynamic
quantities,\foot{Monotonicity of the thermodynamic c-function considered in
\refs{\Appelquist} follows from the inequality $\langle T^\mu_\mu \rangle
\geq 0$, where $\langle\rangle$ indicates a thermal average and
$T_{\mu\nu}$ is the stress tensor in what we would call the boundary
theory.} and \refs{\AlvarezGomez} uses the AdS/CFT correspondence.  When
this work was complete, we learned that an equivalent form of the
c-function in \CFunctionKink\ was discussed in the later versions of
\refs{\GPPZ}.\foot{We thank A.~Zaffaroni for bringing this work to our
attention, and also M.~Porrati for subsequent discussions.}  Its
monotonicity was checked for the flows considered there.

\newsec{Supersymmetric flows}
\seclab\Superkink

To make the correspondence between the supergravity theory and the boundary
theory on the brane more complete, one should be able to find the flow
between the central critical point with maximal supersymmetry and the
$\cN=2$ supersymmetric critical points described in
section~\SuperBackgrounds. In particular one should be able to preserve
$\cN=1$ supersymmetry on the branes {\it all along the flow}, {\it i.e.},
the kink should itself be supersymmetric.  Thus we once again study the
Killing spinor conditions, \ie\ the vanishing of \fermvars, but the metric
now takes the form \RGFmetric\ and the scalars depend on $r$.  The vanishing
of $\delta \chi_{abc}$ directly relates $r$ derivatives of the scalars
to the gradient of $W$. Specifically, to solve $\delta \chi_{abc} =0$ we
find that we must first set $\phi = 0$. There are then two classes of
supersymmetry preserving kink solutions that are related by
$\varphi_1\leftrightarrow \varphi_2$. Here we will describe the family that
contains the flow to \susyGS. We will use the same $\gamma$-matrix
conventions as \GRW.

For an unbroken supersymmetry along the flow of the form
\eqn\unbrknsusy{\epsilon^a  ~=~  \big(
v_1^a ~+~ \tilde v_1^a \gamma^r \big)~e^{A(r)/2} \eta \ ,}
 where $v_1^a$ and $\tilde{v}_1^a$ are given in \USPevecs\ and $\eta$ is a
constant spinor, the vanishing of $\delta \chi_{abc}$ requires:
\eqn\steepdesc{{d \varphi_j \over d r} ~=~
{g \over 2}~{\del W \over \del \varphi_j} \ ,}
where $W= \lambda_1|_{\phi = 0}$ (see \Wevals).
In this section, except as indicated explicitly, the subscript $j$ on
$\varphi_j$ runs from~$1$ to~$3$.
One can show that for such an $\epsilon^a$ to satisfy
the symplectic-Majorana condition in five dimensions,
$\eta$ must be a Majorana spinor
in the four dimensions of the brane.  To verify this one
uses: $\Omega^{ab} v_1^b = -\tilde v_1^a$,
$\Omega^{ab} \tilde v_1^b = v_1^a$, where
$\Omega$ is the $USp(8)$ symplectic form. One also needs
the fact that the four-dimensional charge conjugation matrix,
$\hat C$, may be written as $\hat C =\gamma^r C$, where $C$ is
the five-dimensional charge conjugation matrix.
As a result, we see that the kink preserves a single,
four-dimensional Majorana supersymmetry, \ie\ the flow
preserves $\cN=1$ supersymmetry on the brane.  Mathematica was 
again essential to deduce the simple result \steepdesc\ from the
$48$ conditions $\delta\chi_{abc} = 0$.

The vanishing of $v_1^a \delta \psi_{\mu a}$ and
$\tilde v_1^a \delta \psi_{\mu a}$ yields one more
equation:
\eqn\AWreln{A' ~=~ - {g \over 3}~W \ .}
 One can easily verify that a solution to the first order equations
\steepdesc\ and \AWreln\ automatically satisfies both the gravitational and
scalar equations of motion arising from \SGAction, \scalarkin\ and \VfromW.

The other family of supersymmetric kinks is obtained by
taking $W= \lambda_2|_{\phi = 0}$, and replacing $v_1,
\tilde v_1$ by $v_2,\tilde v_2$.  The conclusions are otherwise
identical.

As promised in section~\CTheorem, we have produced a function of the
scalars, $W$, which is related to the c-function by
\eqn\CFunctionAgain{
   {\cal C}(r) = {27 \over 32} {N^2 \over |W|^3} \ ,}
and whose gradient flow trajectories determine the scalar profiles in
the supergravity kink. Since the scalars are viewed in the AdS/CFT
correspondence essentially as parameters in the lagrangian, we can
assert that the supersymmetric renormalization group flow is gradient
flow. The monotonicity of $A'$ is of course a trivial consequence:
using \AWreln\ and \steepdesc\ one obtains
\eqn\newmonot{A'' ~=~ - {g^2 \over 6}~\sum_j~ \Big| {\del W \over
\del \varphi_j} \Big|^2~=~ - {2 \over 3}~\sum_j~ \Big| {\del
 \varphi_j \over \del r } \Big|^2 \leq 0 \ .}
The monotonicity of $A'$ is related to the local potential energy of
the superkink. Perhaps more fundamentally, recall that the value of the
superpotential at either end of a kink may be thought of as determining
the topological sector \DOEW.\foot{Unlike standard
kinks, where a field flows from one minimum of a potential to another,
our kink flows from a local maximum of the gauged supergravity
potential to a saddle point. As long as there are relevant
perturbations in the gauge theory, there will always be directions in
the potential where the second derivative is negative, but that does
not spoil stability of the saddle point vacuum in supergravity provided
the Breitenlohner-Freedman bound is satisfied. Instead of using finite
total energy to restrict the asymptotic behavior of the kink, we can
use the requirement that there be no naked singularities. With this
criterion our kink is locally unique.} Thus the change in ${\cal C}(r)$
as $r$ goes from large positive values (the ultraviolet) to large
negative values (the infrared) is a measure of the topological charge
of the superkink.

Note there are infinitely many supersymmetry preserving solutions starting
at the point of maximal supersymmetry, $\varphi_j = 0, \phi = 0$. Most of
these solutions go to $W \to -\infty$, and it requires the correct initial
conditions to reach the non-trivial supersymmetric critical point.  We can
set $\phi = 0$ and $\varphi_2 = 0$ throughout the flow: the first
restriction is necessary to preserve supersymmetry, and the second is
necessary to preserve the $U(1)$ symmetry which is present at the infrared
fixed point (as commented in section~\FieldTheory, a non-anomalous $U(1)$
current exists for the entire field RG flow, and it should be reflected in
a $U(1)$ symmetry of the supergravity solution).  One can check that ${\del
W \over \del \varphi_2} \big|_{ \varphi_2 = 0}$ vanishes, so $\varphi_2=0$
is also consistent with supersymmetry.  The superpotential now reduces to
\eqn\Wreduced{W~=~ {1 \over 4 \rho^2}~ \Big[\cosh(2 \varphi_1)~
( \rho^{6}~-~ 2)~ - ( 3\rho^{6} ~+~ 2 ) \Big] \ .}
 The contour maps of $V$ and $W$ on the $(\alpha= {1 \over \sqrt{6}}
\varphi_3, \varphi_1)$ parameter space are shown in Figure~3.  The map of
$V$ shows five extrema.  Point $1$ is the maximally supersymmetric point.
It is a local maximum of both $V$ and $W$.  Points $2$ and $3$ are $\ZZ_2$
equivalent $SU(3)$ invariant points.  In the $\varphi_1$ direction these
points sit at a quadratic minimum of $V$; in the $\varphi_3$ direction they
sit on a cubic inflection point.  Points $4$ and $5$ are $\ZZ_2$ equivalent
${\cal N}=2$ supersymmetric points.  These points are generic quadratic
saddle points of both $V$ and $W$, and the contours surrounding them are
locally hyperbolas.  The superkink which we study follows a path from the
central maximum of $W$ down the ``ridge'' to the saddle point labelled $4$.
We have been unable to find an analytic solution to the steepest descent
equations, but we will discuss a numerical solution as well as analytic
treatments of the endpoints at the end of this section.

\goodbreak\midinsert
\vskip .5cm
\centerline{ {\epsfxsize 6cm\epsfbox{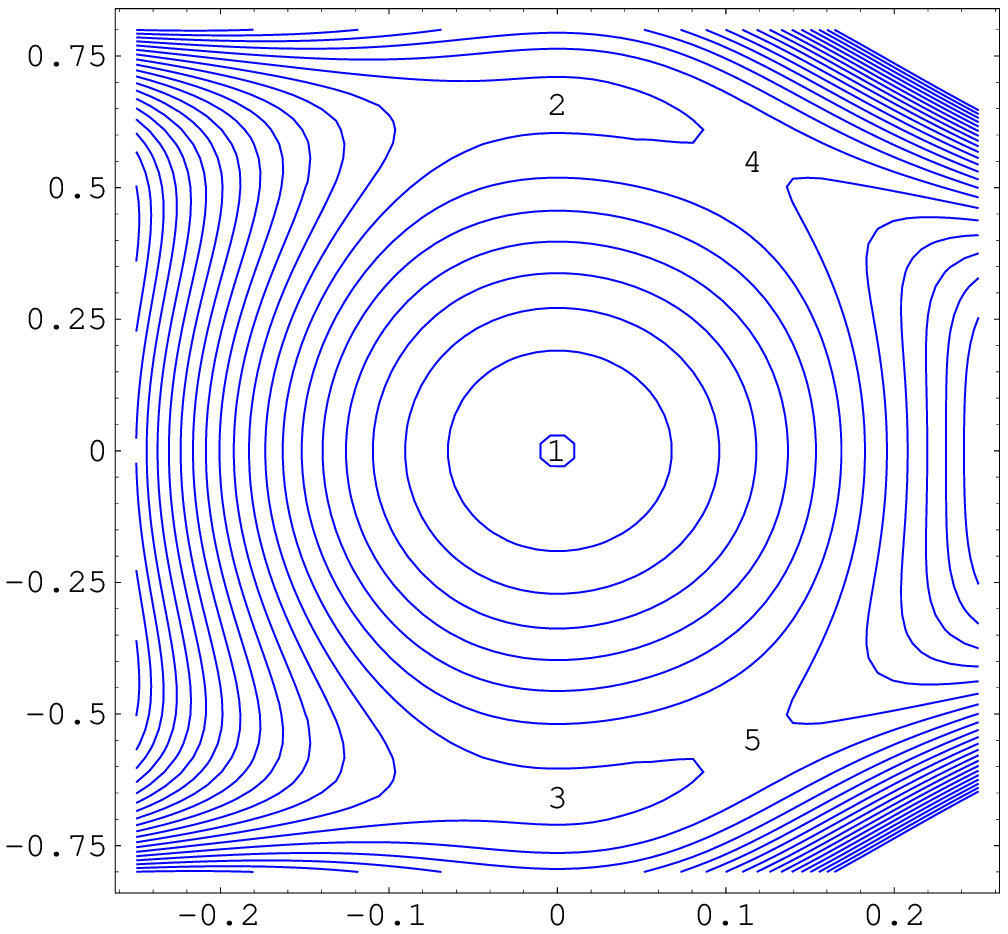}}
\hskip 0.5cm {\epsfxsize 6cm\epsfbox{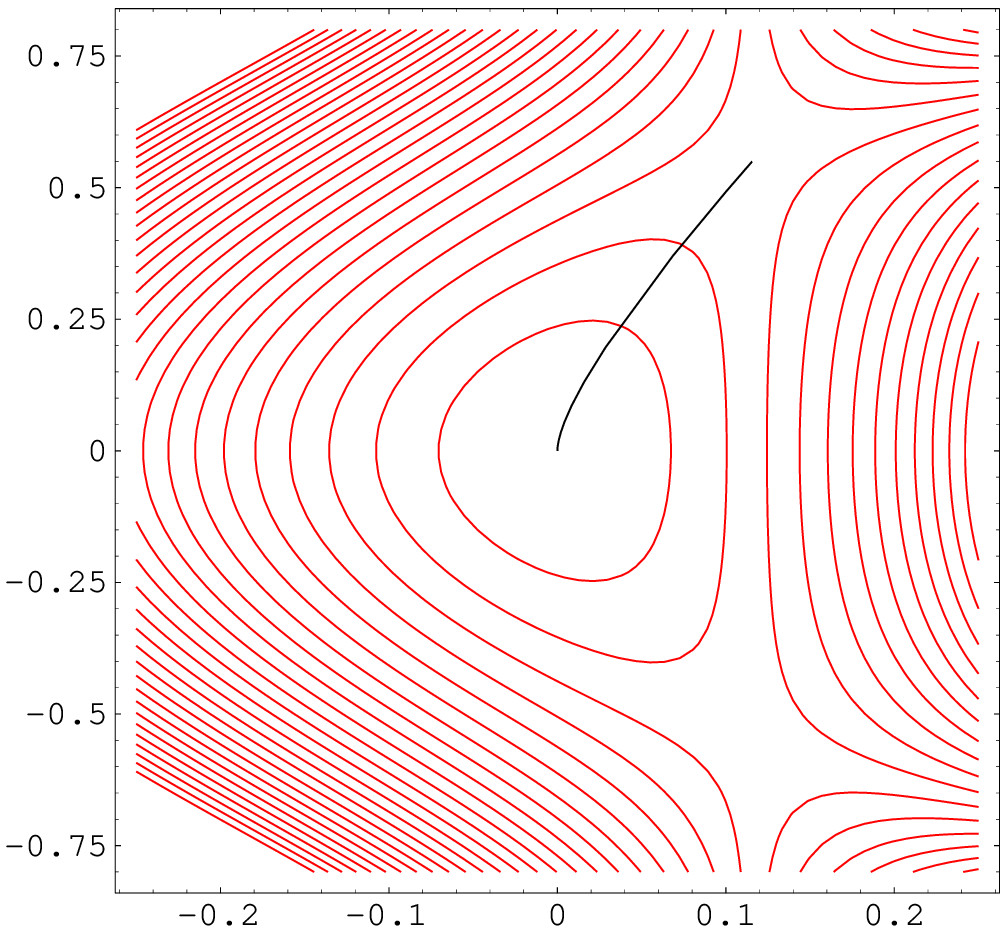}} } \leftskip 2pc
\rightskip 2pc\noindent{\ninepoint\sl \baselineskip=8pt {\bf Fig.~3}:
 The contour map of $V$ (on the left) and $W$ (on the right), with
$\varphi_1$ on the vertical axis and $\alpha = {1 \over \sqrt{6}} \varphi_3$
on the horizontal axis.  The five labelled points are the only extrema of
$V$ in this plane.  $V$ has vanishing first derivatives in all directions
orthogonal to the plane.  A numerical solution of the steepest descent
equations is shown superimposed on the contour plot of $W$.}
 \endinsert

Consider the mass matrix, $M_{ij}$, for the $\varphi_j$ at a critical
point of $W$.  Differentiating \VfromW, setting $\phi=0$ and
using  $\del W/\del \varphi_j  =0$, one gets:

\eqn\hessianV{\eqalign{M_{ij} ~=~ &
{\del^2 V \over \del \varphi_i \del \varphi_j} ~=~
{g^2 \over 4}~ \sum_{k = 1}^3 ~ {\del^2 W  \over \del \varphi_i
\del \varphi_k} ~ {\del^2 W  \over \del \varphi_k \del \varphi_j}  ~-~
{2 g^2 \over 3}~W~{\del^2 W \over \del \varphi_i \del \varphi_j} \cr
{}~=~ & \Big( {g W \over 3}\Big)^2~ \cU_{ik}~ \big( \cU_{kj}~-~ 4
\delta_{kj} \big) \ , \qquad {\rm where} \quad
\cU_{ij}~\equiv~ \Big( {3 \over 2 W}\Big) {\del^2 W  \over \del
\varphi_i \del \varphi_j}\ .}}
The mass scale is set by the inverse radius, $1/\ell$, of the
AdS space, and this may be written $1/\ell = \sqrt{-V/3} =
-{g \over 3} W$, where we have used \VfromW.  For $\phi = 0$ the
matrix $\cU_{ij}$ is real and symmetric, and so has real
eigenvalues, $\delta_k$, $k=1,2,3$.  It follows
that the eigenvalues of $\ell^2 M^2_{ij}$ are simply:
\eqn\CdimsW{m^2_k ~=~ \delta_k(\delta_k ~-~ 4) \ .}
 Combining this with the results of \refs{\SGIKAP,\WitHolOne} we see that a
particular eigenvalue $\delta_k$ of $\big( {3 \over 2 W}\big) {\del^2 W
\over \del \varphi_i \del \varphi_j}$ is related to the conformal dimension
$\Delta_k$ of the field theory operator dual to the variation of the fields
$\varphi_j$ in the corresponding eigen-direction either by $\Delta_k =
\delta_k$ or $\Delta_k = 4 - \delta_k$.

Since $U = e^{A(r)}$ is the renormalization scale on the flow, we
should be able to read off the leading contributions to the
$\beta$-functions of the couplings $\varphi_j$ in the neighborhood of
the end points of the flow. Note that ${d \over dr} = A'~U {d \over d
U} = -{g \over 3} W~ U {d \over d U}$. Hence, \steepdesc\ becomes
\eqn\betafn{U {d \over d U} ~\varphi_j ~=~ -{3 \over 2}~{1 \over W}~
{\del W \over \del \varphi_j} ~\approx- \cU_{jk} \Big|_{crit. pt.} ~
\delta \varphi_k   \ ,}
where we have expanded to first order in the neighborhood of a critical
point, and where $\cU_{jk}$ is defined in \hessianV.\foot{We regard $U =
e^{A(r)}$ as a natural measure of energy scale because of the conformal
factor $e^{2 A(r)}$ in the metric \RGFmetric.  Other definitions have been
proposed \refs{\PeetPolch}.  For instance, $U$ could be taken as the
inverse of the coordinate time $t$ it takes for a light ray to propagate to
a given radius.  This gives $U = \left( \int_r^\infty d\tilde{r}
e^{-A(\tilde{r})} \right)^{-1}$.  The difference is immaterial in \betafn\
because we are operating close to the UV fixed point.}  The eigenvalues,
$\delta_j$, of $\cU_{jk}$ thus determine the behavior of the scalars
$\varphi_j$ near the critical points.  First, to depart the UV fixed point
($U = +\infty$) the flow must take place in directions in which the
eigenvalues are positive (\ie\ the corresponding operators must be
relevant), and to approach the IR fixed point ($U \to 0$) the eigenvalues
must be negative (\ie\ the corresponding operators must be irrelevant).  
At the central critical point, the eigenvalues are $\delta_k = 1,2,3$,
corresponding to $\varphi_1,\varphi_3,\varphi_2$.  Since $\varphi_2 \equiv
0$ on this flow, only the eigenspaces of eigenvalue $1$ and $2$ are used.
The supergravity fields $\varphi_1$ and $\varphi_3$ are dual to operators
of dimension $3$ and $2$, respectively: the fermion and scalar mass terms.

Indeed, one can perform the analysis of the ridge-line flow
in the neighborhood of $\varphi_i = 0$ by considering the series expansion 
  \eqn\WSeries{
   W = -{3 \over 2} - {\varphi_1^2 \over 2} - \varphi_3^2 + 
    \sqrt{8 \over 3}~\varphi_1^2 \varphi_3 + \ldots \ ,
  }  
 where we have set $\varphi_2 = 0$.  We will operate in units where $g=2$.
Keeping only the quadratic terms in the expansion leads to $\varphi_1 \sim
\pm e^{-r}$, $\varphi_3 \sim \pm e^{-2r}$, and with this behavior the cubic
term is of the same order as the quadratic terms.  We must keep this term
even in a lowest-order analysis of the behavior of $\varphi_3$.  The
correct analysis of the asymptotics is
\eqn\UVAsympt{
\eqalign{
\left[ {d \over dr} + 1 \right] \varphi_1 &= 0  \cr
\left[ {d \over dr} + 2 \right] \varphi_3 &= \sqrt{8 \over 3}~
\varphi_1^2} \qquad\quad
\left. \eqalign{
\varphi_1 &\sim a_0 e^{-r}  \cr
\varphi_3 &\sim \sqrt{8 \over 3}~a_0^2~r e^{-2r} + a_1 e^{-2r} } \ 
\right\}
\quad \hbox{as $r \to \infty\,$.}}
 It is straightforward to verify that higher order terms in \WSeries\ do
not change this story: roughly speaking, two powers of $\varphi_1$ are
worth one power of $\varphi_3$ in the expansion of $W$.  The integration
constant $a_0$ is, up to factors of order unity, the fermion mass resulting
from the term ${1 \over 2} m \tr \Phi_3^2$ in the superpotential (see
\SuperPot).  Supersymmetry fixes the value of the coefficient of $r
e^{-2r}$ in the asymptotics of $\varphi_3$.  This coefficient, again up to
factors of order unity, is the square of the boson mass resulting from the
mass term in the superpotential.  The integration constant $a_1$ is not
determined by supersymmetry, and reflects the possibility of different
states in the theory, only one of which is the vacuum which preserves
conformal invariance.  We do not yet have a way of determining $a_1$
analytically, but we will shortly discuss its numerical evaluation.

The flow near the IR critical point \susyGS\ is a little simpler.  The
eigenvalues, $\delta_k$, of $\cU_{ij}$ are $1+\sqrt{7},\, 1-\sqrt{7},\, 3$.
The only negative one is $1-\sqrt{7}$, and so the corresponding eigenvector
determines the direction from which the flow approaches the fixed point.
The asymptotics (again in units where $g=2$) is 
  \eqn\IRAsympt{\seqalign{\span\TC}{
   \left. \eqalign{\varphi_1 &\sim {1 \over 2} \log 3 - 
    b_0 e^{\lambda r}  \cr\noalign{\vskip-0.5\jot}
    \varphi_3 &\sim {1 \over \sqrt{6}} \log 2 - 
    {\sqrt{7}-1 \over \sqrt{6}} b_0 
    e^{\lambda r}
   } \ \right\} \quad \hbox{as $r \to -\infty$.}  \cr\noalign{\vskip1\jot}
   \lambda = {2^{5/3} \over 3} (\sqrt{7}-1)}
  }
 The irrelevant operator in the field theory that controls this flow has
dimension $3+\sqrt{7}$.  Because of the possibility of shifting $r$
additively, the coefficients $a_0$, $a_1$, and $b_0$ are not all
independently meaningful.  The two combinations that are invariant are $b_0
a_0^\lambda$ and ${a_1 \over a_0^2} + \sqrt{8 \over 3} \log a_0$.  Additive
shifts of $r$ amount to rigid dilations of the boundary, and the flow is
invariant under these if we make a corresponding shift of the parameter $m$
in the superpotential.

A numerical solution interpolating between \UVAsympt\ and \IRAsympt\ is
easier to find if one starts from the infrared end.  Then the differential
equations are stable in that a small perturbation in the initial conditions
damps out over time.  The opposite is true of the flow from the ultraviolet
end to the infrared end, which is appropriate since relevant deformations
become more important in the infrared.  Our results for the invariant
combinations of coefficients are
  \eqn\CoefNumbers{
   b_0 a_0^\lambda \approx 0.1493 \qquad\quad
   {a_1 \over a_0^2} + \sqrt{8 \over 3} \log a_0 \approx -1.4694 \ .
  }
 Minus the logarithm of the first number can be thought of roughly as the
width of the kink.  The second number represents the choice of vacuum state
that leads to infrared physics with non-vanishing central charge.

\newsec{Supergravity fields and the operator map}
\seclab\sugraandopmap

The symmetries of the supersymmetric critical point \refs{\KPW} in $AdS_5$
include ${\cal N}=2$ supersymmetry and the gauge symmetry $SU(2)_I$, so the
supergravity fields should be classified in representations of the
corresponding superalgebra $SU(2,2|1)\times SU(2)_I$. The AdS/CFT
correspondence requires a one to one map between these fields and gauge
invariant composite operators of the boundary theory, in this case the mass
deformed ${\cal N}=4$ super-Yang-Mills theory discussed in
section~\FieldTheory.  In ${\cal N}=1$ superspace this theory contains a
chiral spinor superfield $W_{\alpha}$ describing the gauge multiplet, and
three chiral scalar superfields $\Phi_i$, $i=1,2,3$, describing
supermatter.  The component
expansions are 
\eqn\CompExp{\eqalign{
W_{\alpha} & = \lambda_\alpha +
\theta^\beta\left[\half(\sigma^\mu\overline\sigma{}^\nu)_{\beta\alpha}
F_{\mu\nu}+i\epsilon_{\alpha\beta}D\right]-
\theta^2i\sigma^\mu_{\alpha\dot\beta}D_\mu\overline
\lambda{}^{\dot\beta} \,, \cr
\Phi_i & = z_i + \theta^{\alpha} \psi_{\alpha i} + \theta^2 F_i\,.\cr}}

In this section we will discuss the spectrum of fields in bulk
supergravity theory and propose a map to operators of the boundary
theory.  Some aspects of the corresponding unitary representations of
$SU(2,2|1)$ \refs{\FlatFrons,\DobrPetk} are summarized in Appendix~B
with a brief discussion at the end of this section.

The main technical task is to study the action of the gauged
supergravity theory of \refs{\GRW,\PPvN} expanded to quadratic order
about the ${\cal N}=2$ critical point and to compute the mass
eigenvalues of all fields. This is an unfortunately complex matter. The
general approach is discussed in Appendix~A, but details reside in the
privacy of our Mathematica notebooks, which shall remain sealed until
the statute of limitations expires. We will give the decompositions of
supergravity fields -- which transform under $USp(8)\times SU(4)$ in
the full theory -- with respect to the unbroken gauge subgroup
$SU(2)_I\times U(1)_R$ at the critical point. This will help readers
understand the group-theoretic structure of the results. The detailed
masses, however, depend crucially upon accurate calculation of the
scalar 27-bein ${\cal V}_{ab}{}^{AB}$ at the critical point and of the
$USp(8)$ tensors $W_{ab}$ and $A_{abcd}$, which are given by a
quadratic and a quartic expression, respectively, in terms of that
27-bein. Moreover, the computation of the masses of scalar fields
requires quadratic expansions of those tensors about the critical
point, and a similarly complicated determination of the kinetic term.

\subsec{The fermion masses}

The relevant part of the supergravity action reads \GRW:%
\foot{In this section, to conform with the notation in
\refs{\GRW} we use indices $\mu,\nu$ for space-time indices in five
dimensions.}
\eqn\FerAct{\eqalign{
e^{-1}L_F= & -\half i \overline \psi{}^a_\mu\gamma^{\mu\nu\rho}
D_\nu\psi_{\rho a}+\hbox{${1\over 12}$}
i\overline\chi{}^{abc}\gamma^\mu D_\mu\chi_{abc} -
\hbox{${1\over 4}$} i g W_{ab} \overline
\psi{}^a_\mu\gamma^{\mu\nu}\psi{}^b_\nu
\cr & +\hbox{${1\over 6}\sqrt{1\over 2}$} ig
A_{dabc}\overline\chi{}^{abc}
\gamma^\mu\psi_\mu^d
+ \half ig \overline\chi{}^{abc}(\half A_{bcde}-\hbox{${1\over 12}$}
\Omega_{bd}W_{ce})\chi_a{}^{de}\,,
\cr} }
where the spin-3/2 field $\psi^a_\mu$ and the spin-1/2 field
$\chi^{abc}$ transform as $\bf 8$ and $\bf 48$ of $USp(8)$,
respectively, and satisfy the symplectic Majorana conditions.

There is a super-Higgs effect in this sector. The simplest way to
treat it is to observe that the spin-1/2 supersymmetry transformation
rule in \refs{\GRW} reduces  to, cf. \fermvars,
\eqn\DelChi{
\delta\chi_{abc} =-\hbox{$\sqrt{1\over 2}$} g A_{dabc} \epsilon^d\,,
}
and this describes a rank 6 linear transformation from the
8-dimensional space of the $\epsilon^a$ into the 48-dimensional space
of the $\chi_{abc}$. Thus one can fix the gauge by using 6 spinor
parameters of the broken supersymmetries to eliminate the 6 fields in
the range of the transformation. Those are precisely the same 6 fields
as in the mass-mixing term in the Lagrangian \FerAct\ which can thus be
dropped in this gauge. The action on
the space of remaining fields has no mixing between the spin-3/2 and
spin-1/2 fields and it is clear that the remaining spin-3/2 part of
the Lagrangian is not modified in the process.

The final diagonalization of the spin-3/2 action can be described quite
explicitly. Since the kinetic term is already in the canonical form,
all that remains is to diagonalize the symmetric matrix $W_{ab}$, which
at the critical point is given in (A.42). It has two 2-dimensional
eigenspaces and one 4-dimensional eigenspace as follows
$$
\seqalign{
\span\TR\qquad & \span\TR\qquad & \span\TR\qquad}
{\lambda_1 = - 2^{2/3}\,, &   v_1=(1,0,0,0,0,1,0,0)\,, &
        \tilde v_1=(0,1,0,0,-1,0,0,0)\,, \cr
\lambda_2= -{4 \over 3} 2^{2/3}\,, &
 v_2=(0,0,0,1,0,0,1,0)\,, &  \tilde v_2=(0,0,1,0,0,0,0,-1)\,, \cr
\lambda_3= -{7 \over 6} 2^{2/3}\,, & v_3=(0,0,1,0,0,0,0,1)\,, &
         \tilde v_3=(0,0,0,1,0,0,-1,0)\,, \cr
\lambda_4= -{7 \over 6} 2^{2/3}\,,&  v_4=(0,1,0,0,1,0,0,0)\,, &
         \tilde v_4=(1,0,0,0,0,-1,0,0)\,.}
$$
Define projections $\psi_{\mu,i}^1=\psi_\mu^a v_{ia}$,
$\psi_{\mu,i}^2=\psi_\mu^a \tilde v_{ia}$ of $\psi_\mu^a$ onto the
eigenspace $(v_i,\tilde v_i)$. In terms of those modes we get a
standard kinetic term, with the symplectic matrix $\omega$ acting by
$\omega v_i=\tilde v_i$, $\omega \tilde v_i=-v_i$, and a standard
symplectic mass term
\refs{\Shuster}. The quadratic action in each eigenspace reads
\eqn\GravMass{e^{-1}L_{3/2}={1\over 2}
\omega_{\al\be}\overline\psi{}_{\mu,i}^\alpha\gamma^{\mu\nu\rho}D_\nu
\psi{}_{\rho,i}^\beta
+
{1\over  2}{m_i\over \ell}\delta_{\alpha\beta}
\overline\psi{}_{\mu,i}^\alpha
\gamma^{\mu\nu} \psi{}_{\nu,i}^\beta\,,}
where
\eqn\Mstuff{
{m_i\over\ell}={3\over 2}g\lambda_i\,,\qquad {1\over \ell} = -{1\over
3}gW = {2^{2/3}\over 3} g\,.}
As expected the sector with
$m_1=(3/2)g\ell \lambda_1$ is the gravitino of the ${\cal N}=2$ theory
and the remaining states are massive spin-3/2 fields with masses
$m_i$, $i=2,3,4$, given in \Mstuff.  This is consistent with the
decomposition $USp(8) \rightarrow SU(2)_I\times U(1)_R$,
\eqn\gravdecom{
{\bf 8}\quad \rightarrow\quad ({\bf 2}_{1/2}\oplus {\bf
2}_{-1/2})\oplus
({\bf 1}_{1}\oplus{\bf 1}_{-1})\oplus 2\times {\bf 1}_{0}\,. }
of which the first two singlets, $\r11\oplus\r1{-1}$, correspond to
the gravitino in ${\cal N}=2$ theory.

The diagonalization of the action for the remaining $42$ spin-1/2
fields is more involved and we have carried it out in Mathematica
using formulae from Appendix~A. As with spin-3/2, the
symplectic Majorana condition requires that the masses occur in pairs
$\pm m$. We have checked that the mass eigenstates transform correctly
under $SU(2)_I\times U(1)_R$ as predicted by the following branching
rule
\eqn\spinbranch{\eqalign{
{\bf 48} \quad \rightarrow  \quad & (\r31\oplus\r3{-1})\oplus
2\times \r30 \oplus 2\times (\r2{3/2}\oplus \r2{-3/2}) \oplus
4\times (\r2{1/2}\oplus \r2{-1/2}) \oplus \cr
& (\r12\oplus \r1{-2})\oplus 3\times (\r11\oplus\r1{-1})\oplus
4\times \r10\,.
\cr}}
The six Goldstino modes that are absorbed into massive spin-3/2 fields
are identified using \gravdecom\ as $(\r2{1/2}\oplus \r2{-1/2})\oplus
2\times \r10$.

\subsec{The boson masses}

The computation of the masses of the vector fields, $A_\mu^{IJ}$, and
the antisymmetric tensor fields, $B_{\mu\nu}^{I\alpha}$, is quite
straightforward and similar to the one above for the fermions. One
must consider two terms in the action
\eqn\VTAct{
e^{-1}L_{AB}=
-\hbox{$1\over 8$} H_{\mu\nu ab}H^{\mu\nu ab} +{1\over 8ge}
\epsilon^{\mu\nu\rho\sigma\tau}\epsilon_{\alpha\beta}\delta_{IJ}
B_{\mu\nu}{}^{I\alpha} D_\rho B_{\sigma\tau}{}^{J\beta}\,,
}
where
\eqn\ThedefofH{
H_{\mu\nu}{}^{ab}=F_{\mu\nu IJ}{\cal V}^{IJab}+B_{\mu\nu}{}^{I\alpha}
{\cal V}_{I\alpha}{}^{ab}\,,
}
and $F_{\mu\nu IJ}$ are the field strengths for the vector fields.
Note that the first term in \VTAct\ gives rise to the kinetic term for
the vectors, but the mass term for the antisymmetric
tensors,\foot{This observation may be rather puzzling at the first
sight -- it is, however, quite natural since the antisymmetric tensors
correspond to field strengths of vector fields in the ungauged theory
\refs{\GRWplb,\GRW,\PPvN}.} and is the only source of cross-terms.  At the
quadratic level the Chern-Simons like kinetic term for the
antisymmetric tensors is invariant under a shift
\eqn\AsTsh{
B_{\mu\nu}{}^{I\alpha}\quad \rightarrow \quad B_{\mu\nu}{}^{I\alpha}
+X^{I\alpha}{}_{JK}F_{\mu\nu}{}^{JK}\,,
}
in which  $X^{I\alpha}{}_{JK}$ is chosen to remove the cross terms.
In this process the
kinetic term and thus the mass eigenvalues of the vector fields are
modified.

The decomposition of the vector fields with respect to $SU(4)\supset
SU(2)_I\times U(1)_R$ is
\eqn\VectDec{
{\bf 15}\quad \rightarrow \quad
\r30\oplus \r10\oplus (\r2{3/2}\oplus\r2{-3/2})\oplus
(\r2{1/2}\oplus \r2{-1/2}) \oplus (\r11\oplus\r1{-1})\oplus \r10\,.
}
One expects that all but the first two modes, corresponding to
unbroken gauge symmetries, become massive. As we have discussed above,
the
actual values of the masses are effected by the mixing with the
antisymmetric tensor fields. However, the usual Higgs effect, in which
11 massless Goldstone bosons are eaten by the massive vector fields,
does not directly affect the calculation. Therefore the masses can be
determined
from \VTAct\ once the proper shift \AsTsh\ has been made.

Since the antisymmetric tensor fields have a first order action, their
masses come in $\pm m$ pairs corresponding to the terms in the
$SU(2)_I\times U(1)_R$ decomposition
\eqn\ATdec{
{\bf 12}\quad\rightarrow \quad 2\times (\r2{1/2}\oplus \r2{-1/2})\oplus
2\times (\r11\oplus \r1{-1})\,.
}
As for the vectors, the main difficulty here is to compute the metric
$({\cal V}{\cal V})_{I\alpha\, J\beta}$ that encodes the dependence of
the masses on the expectation values of the scalar fields.

The calculation of masses of the scalar fields presents additional
difficulty if one wants to consider the full set of 42 fields, rather
than the restricted subset corresponding to the gauge fixed subspace
of the scalar manifold $\cal S$ discussed in section~\SugraandYM.
In the general
case one must calculate carefully both the $\sigma$-model kinetic
action at the critical point and expand the potential to the quadratic
order in all directions.  The calculation can be simplified by
restricting to different sectors of $SU(2)_I\times U(1)_R$, one at a
time, which in this case are given by the branching
\eqn\Scalbr{\eqalign{
{\bf 42}\quad \rightarrow\quad &
\r30\oplus 2\times (\r31\oplus \r3{-1})\oplus
(\r2{3/2}\oplus \r2{-3/2})\oplus 3\times (\r2{1/2}\oplus
\r2{-1/2})\oplus
\cr &
2\times (\r12\oplus \r1{-2}) \oplus (\r11\oplus \r1{-1}) \oplus
5\times \r10\,.\cr
}}
By carrying out a careful diagonalization we find a perfect agreement
of the mass spectrum with this decomposition. In particular, the
result displays 13 massless states of which the two real $SU(2)_I$
singlets are the dilaton and the axion and the remaining 11 are the
Goldstone bosons from the breaking $SU(4)\rightarrow SU(2)_I\times
U(1)_R$.  Their quantum numbers are in agreement with those of massive
vectors in \VectDec.

\subsec{The $SU(2,2|1)$ spectrum and the operator map}

The ultimate test of our result for the mass spectrum is to verify
that it can be assembled into irreducible unitary representations of
the superalgebra $SU(2,2|1)\times SU(2)_I$.

The relevant representations of $SU(2,2|1)$ are parametrized by the
energy $E_0=\Delta$, of the lowest weight state, equal to the scaling
dimension of the corresponding boundary operator, the two spins $j_1$
and $j_2$ and the $R$-charge, $r$. These quantum numbers designate a
representation of the bosonic subalgebra $U(1)\times SU(2)\times
SU(2)\times U(1)$ where the first and last $U(1)$ factors are the
energy and $R$-symmetry, respectively. The relation between $\Delta$
and the mass for various fields may, for example, be found in
\FFZ\ for $d=4$. In general, for ${\rm AdS}_{d+1}$, the relations are
\smallskip
\item{(i)}
scalars \WitHolOne:\quad $\Delta_{\pm} = \half(d \pm \sqrt{d^2
+4m^2})$,
\item{(ii)} spinors \refs{\HSspin}:\quad  $\Delta = \half (d + 2|m|)$,
\item{(iii)} vectors:\quad
$ \Delta_{\pm} = {1 \over 2} (d \pm \sqrt{(d-2)^2 + 4m^2})$,
\item{(iv)} $p$-forms \refs{\lYi}:\quad
$ \Delta = {1 \over 2} (d \pm \sqrt{(d-2p)^2 + 4m^2})$,
\item{(v)} first-order $(d/2)$-forms ($d$ even):\foot{See \GEASAF\ 
for $d=4$.  Entry (v) is for forms with first order  actions \PvNT, 
while entries (iii) and (iv) are for forms with Maxwell type actions, possibly 
augmented by mass terms.}\quad
$\Delta={1\over 2}(d + 2|m|)$,\comment{Krzysztof: $2|m|$ or $2p$?  What
would $m$ be for a 1st order action?  SSG}
\item{(vi)} spin-3/2  \refs{\AVol,\KosRyt}:\quad
$\Delta = \half(d + 2|m|)$,
\item{(vii)} massless spin-2 \refs{\WitHolOne}:\quad
$\Delta = d$.

We have converted results for mass eigenvalues of all fluctuations
into scale dimensions by the formulas. Results are tabulated
in Tables~6.1 and 6.2 in which each
horizontal band corresponds to a representation of the product of
$SU(2,2|1)$ with the $SU(2)_I$ flavor symmetry. The
detailed structure of these $SU(2,2|1)$ representations is given in
Appendix~B, and the entry in the tables denotes the lowest weight
energy and the $SU(2)_I\times U(1)_R$ quantum numbers at each level.
In all but one case we take $\Delta=\Delta_+$, the largest root given
above. The unique exception is the complex scalar triplet with
$\Delta=\Delta_-=3/2$. This assignment is required by the well known
relation $\Delta = 3r/2$ for ${\cal N}=1$ chiral primary operators
and, of course, by the structure of the $SU(2,2|1)$ representation.
The choice $\Delta=\Delta_-$ was made, also for group theoretic reasons, 
in another recent study
\refs{\samson} of the AdS/CFT correspondence. It is also known that the
chiral multiplet in $AdS_4$ supersymmetry requires quantization with
both ``regular'' and ``irregular'' boundary conditions \refs{\PBDF}
when scalar masses are in the range $-d^2/4 < m^2 < 1 - d^2/4$. See
\refs{\MezTow} for a discussion valid in $AdS_{d+1}$.

There is presently no coherent prescription to calculate correlation
functions of scalar operators of dimension $\Delta < 2$ from the AdS/CFT
correspondence, and the two-point function appears to be particularly
difficult.  The following procedure seems a provisionally satisfactory
application of the prescription of \refs{\SGIKAP,\WitHolOne}.  Consider the
three-point correlator $\langle J_\mu {\cal O}^* {\cal O} \rangle$, where
${\cal O}$ is the operator of "irregular" dimension and $J_\mu$ is a
conserved current of the boundary theory.  In the present example $J_\mu$
could be one of the global $SU(2)$ currents or the R-symmetry current, both
of which correspond to fields in five-dimensional gauged supergravity.
Note that we do not specifically rely on supersymmetry here, but only on
the existence of some conserved current under which ${\cal O}$ is charged.
The three-point function above was studied in \refs{\fmmr}.  The required
integral over $AdS_5$ converges for $\Delta > 1$, exactly the range needed!
The two-point function $\langle {\cal O}^* {\cal O} \rangle$ can be then
obtained from the Ward identity for $J_\mu$, and it has the correct power
law, namely $1/|x-y|^{2\Delta}$.\foot{We thank E.~Witten and L.~Rastelli
for useful discussions on this point.}


\def\ccD#1{{\cal D}({\textstyle #1})}
\def\TCtextstyle{\hfil$\textstyle{##}$\hfil}

\goodbreak\midinsert
$$
\seqalign{
\span\TC\qquad & \span\TCtextstyle\qquad\ \  & \span\TR\qquad\quad &
\span\TR\qquad & \span\TR\qquad\quad & \span\TR\qquad\quad
& \span\TR\qquad & \span\TR }
{& \Delta  &\phi  & \chi & A_\mu &B_{\mu\nu}
                &\psi_\mu& h_{\mu\nu}\cr
\noalign{\medskip\hrule\medskip}
\ccD{{3\over 2},0,0;1} &  {3\over 2}  &{\bf 3}_1 &&&&& \cr
{\rm Tr}\,\Phi_i\Phi_j  &  2     & &\r30&&&& \cr
{\rm complex} &  {5\over 2}   &{\bf 3}_{-1} &&&&& \cr
\noalign{\medskip\hrule\medskip}
\ccD{2,0,0;0} &  2   &{\bf 3}_0 &&&&& \cr
{\rm Tr}\,\overline \Phi T^A\Phi  &  {5\over 2}
                & &\r31\oplus\r3{-1}&&& \cr
{\rm real} &  3   & &&\r30&&&\cr
\noalign{\medskip\hrule\medskip}
\ccD{{9\over 4},\half,0;{3\over 2}}
 &  {9\over 4}   & &\r2{3/2}&&&& \cr
{\rm Tr}\,W_\alpha\Phi_{j}  &  {11\over 4 }   &\r2{1/2} &&&\r2{1/2}&
\cr
{\rm complex} &  {13\over 4}   & &\r2{-1/2}&&&&\cr
\noalign{\medskip\hrule\medskip}
\ccD{3,0,0;2} &  3   &{\bf 1}_2 &&&&& \cr
{\rm Tr}\,W^\alpha W_\alpha  &  {7\over 2}     & &\r11&&&& \cr
{\rm complex} &  4   &{\bf 1}_{0} &&&&& \cr
\noalign{\medskip\hrule\medskip}
\ccD{3,\half,\half;0} & 3  &&&\r10&& \cr
J_{\alpha\dot\alpha}  &   {7\over 2}     & &&&&\r11\oplus\r1{-1}& \cr
{\rm real} &  4   & &&&&& \r10 \cr
\noalign{\medskip\hrule\medskip}
 &     & &&&&&  }
$$
\noindent{\ninepoint\sl \baselineskip=8pt {\bf Table~6.1}:
The five short $SU(2,2|1)$ representations in the mass spectrum of
supergravity fields at the ${\cal N}=2$ critical point
and the corresponding ${\cal N}=1$ superfields in the boundary gauge
theory. For each complex representation above there is an additional
representation with the same dimension, spin and $SU(2)_I$ content, but
opposite $U(1)_R$ charge.  }
\endinsert

\goodbreak\midinsert
$$
\seqalign{
\span\TC\quad & \span\TCtextstyle\qquad & \span\TR\qquad &
\span\TR\qquad &
 \span\TR\qquad & \span\TR\qquad & \span\TR\qquad & \span\TR }
{& \Delta  &\phi  & \chi & A_\mu &B_{\mu\nu}
                &\psi_\mu\cr
\noalign{\medskip\hrule\medskip}
\ccD{\Delta,0,0;0} &  1+\sqrt{7}   &\r10 &&&& \cr
\Delta=1+\sqrt{7} &  {3\over 2}+\sqrt{7}   & &\r11\oplus\r1{-1}&&& \cr
K &  2+\sqrt{7}   &\r12\oplus\r1{-2} &&\r10&& \cr
{\rm real}  &   {5\over 2}+\sqrt{7}   & &\r11\oplus\r1{-1}&&& \cr
  & 3+\sqrt{7}  &\r10 &&&& \cr
\noalign{\medskip\hrule\medskip}
\ccD{{11\over 4},\half,0;\half} &  {11\over 4}   & &\r2{1/2}&&&
\cr
\Lambda_{\alpha i} & {13\over 4}    &\r2{-1/2}
        &&\r2{3/2}&\r2{-1/2}& \cr
{\rm complex} & {15\over 4}    & &\r2{1/2}\oplus\r2{-3/2}&&&\r2{1/2}
\cr
 &  {17\over 4}   & &&\r2{-1/2}&& \cr
\noalign{\medskip\hrule\medskip}
\ccD{3,0,\half;\half} &   3  & &\r10&&& \cr
\Sigma_\alpha
        &  {7\over 2}  & &&\r1{-1}&\r1{1}& \cr
{\rm complex}  &   4  & &\r1{-2}&&&\r10 \cr
 &  {9\over 2}   & &&&\r1{-1}& \cr
\noalign{\medskip\hrule\medskip}
 &     & &&&& }
$$
\noindent{\ninepoint\sl \baselineskip=8pt {\bf Table~6.2}:
The remaining $SU(2,2|1)$ representations in the supergravity mass
spectrum.
The representations labelled complex are again doubled as described in
the
caption for Table~6.1.}
\endinsert

The short representations and corresponding superfields%
\foot{
The relation between $SU(2,2|1)$ representations and boundary
superfields in the context of the AdS/CFT correspondence has recently
been discussed in \refs{\FerrZaff,\FerrLLZaff}.
}
 are listed in
Table~6.1.  They are either chiral, and thus satisfy $\Delta = 3r/2$
for the lowest component, or they have protected dimensions because
the multiplet contains one or more conserved component operators: for
example, the $SU(2)$ flavor current in ${\rm Tr}\,\bar\Phi T^A\Phi$,
and the $U(1)_R$ current, supercurrent, and stress tensor in the
supercurrent superfield $J_{\alpha
\dot\alpha}$.  Since the massive field
$\Phi_3$ is effectively integrated out in the flow, the IR theory
contains the massless chiral superfields $\Phi_i\,,$ $i=1,2$ with
$\Delta=3/4$ and $r=1/2$ and the gauge superfields $W_\alpha$ with
$\Delta=3/2$ and $r=1$.  We see that the short multiplets found from
$AdS_5$ supergravity contain {\it all bilinears} formed from these
fields, all with correct quantum numbers. This is a striking
confirmation of the validity of the correspondence in a new and more
complex theory.

The first long multiplet in Table~6.2 is a massive vector
representation
of $SU(2,2|1)$ with irrational and clearly unprotected $\Delta$. We
use the designation $K(x,\theta,\bar \theta)$ to indicate that it
corresponds to a general scalar superfield in the boundary
theory.

The existence of long multiplets of $SU(2,2|1)$ in the supergravity
spectrum should not come as a surprise. In fact, in \refs{\gEin} it was
suggested that for the Kaluza-Klein compactification of ten-dimensional
type IIB supergravity on the $T^{11}$ coset manifold (which again leads
to a supergravity geometry with $SU(2,2|1)$ symmetry), only a small
fraction of operators will have rational dimensions. The fraction is on
the order $1/\Delta^2$ for operators of dimension less than or equal to
some large maximum, $\Delta$. In supergravity, the many fields of
irrational but finite dimension arise from a Kaluza-Klein
compactification on a manifold of smaller isometry than $S^5$. In field
theory, the natural interpretation is simply to say that the smaller
$SU(2,2|1)$ algebra protects fewer operators than $SU(2,2|4)$. The
outstanding puzzle is why the operators dual to supergravity fields in
long multiplets should retain finite dimensions while the operators
dual to excited string states acquire dimensions on the order
$(g_{YM}^2 N)^{1/4}$.

We would like to identify the scalar superfield
$K(x,\theta,\bar\theta)$ with an operator in the gauge theory. We
speculate that it is the K\"ahler potential.  This is a sensible guess
because the K\"ahler potential evolves in the RG flow, and the scalar
field that measures the approach of our trajectory to the ${\cal N}=1$
point sits in the supergravity multiplet dual to
$K(x,\theta,\bar\theta)$.  This scalar field has a dimension
$3+\sqrt{7}$ in the infrared.

The lowest weight states of the remaining long multiplets have spin
$1/2$ and maximum spin $3/2$. We assign them to boundary
operators which are non-chiral spinor superfields $\Lambda_{\alpha
i}(x,\theta,\bar\theta)$ and $\Sigma_{\alpha}(x,\theta,\bar\theta)$.
The natural candidates for these operators are the 3 supercurrents of
the ${\cal N}=4$ theory which are explicitly broken by the mass term
in (2.1). In the absence of the mass term the 3 extended supercurrents
are components of the ${\cal N}=1$ $SU(3)$ triplet superfield\foot{We
thank  M.~Grisaru for this information.}
\eqn\Supcur{S^i_{\alpha}(x,\theta,\bar\theta)
 = {\rm Tr}(W_{\alpha} \overline\Phi^i
+\half \epsilon^{ijk}\Phi_j D_{\alpha} \Phi_k)\,,}
where $D_\alpha$ is the gauge and supercovariant derivative. We suggest
that $\Lambda_{\alpha i}$ and $\Sigma_{\alpha}$ are the $i=1,2$ and
$i=3$ components of this superfield, respectively, and we note that the
R-charges agree with this assignment. The scale dimensions $\Delta$
generally disagree with the naive sum of those of the elementary fields
as is expected for non-chiral operators.

The two massive spin-3/2 representations have strikingly
different component structure, for example $\Lambda_{\alpha i}$
contains a scalar component, but $\Sigma_{\alpha}$ does not. We have
confirmed in Appendix~B that these representations are characterized
by null or shortening conditions which eliminate states, specifically
the condition $2E_0=4(j_2+1)+3r$ with $j_2=0$ for $\Sigma_{\alpha}$
and $2E_0=4(j_2+1)+3r$ with $j_2=\half$ for $\Lambda_{\alpha i}$. Thus
they have a protected structure in the same sense in which a chiral
superfield with $2E_0=3r$ is protected. The structure of these
representations suggests they should not be
called long representations, but perhaps semi-long. The detailed
correspondence between the supergravity fields which are components of
these representations and the components of the superfields
$S^i_\alpha$ is left here as a curious open question.

Although the detailed analysis of the representations of Table~6.2 is
given in Appendix~B, it is worth remarking here that their structures
can be described in terms of the Higgs effect in which a massless
multiplet of $SU(2,2|1)$, which contains a component gauge field, eats
a lower spin multiplet. The massless representations are listed in
Table~5 of \GRW. The states of the massive vector multiplet we have
called $K$ in Table~6.2 are exactly those of a massless vector multiplet
plus a hypermultiplet\foot{A hypermultiplet is the direct sum of a LH
and RH spinor multiplet in the notation of \GRW.}  in which one of the
4 real scalars is eaten. A massive spin-3/2 field has the choice of
eating a spin-1/2 field from a vector or antisymmetric tensor
multiplet. The states of the $\Sigma_{\alpha}$ representation are
those of a massless gravitino multiplet plus a tensor multiplet with
one spin-1/2 field and one scalar eaten to make its massive higher spin
components. On the other hand the states of the $\Lambda_{\alpha i}$
representation are those of a gravitino plus vector plus LH spinor
multiplet with a spinor and two scalars eaten. We have already seen
that no Higgs mechanism is required to make an antisymmetric tensor
massive.

\newsec{${\cal N}=4$ supergravity}
\seclab\NFourSUGRA

Thus far, our explicit analysis of the $SU(2)_I$ invariant
subsector of gauged $\cN=8$ supergravity has focussed largely
upon the scalar manifold.  We now return to the consideration
of the $\cN=4$ supergravity theory associated with the $SU(2)_I$
invariant subsector of the complete $\cN=8$ theory.

The $\cN=4$ supergravity in five dimensions was first considered in
\ECrem, but was only really explicitly described in \MAPT. The latter
reference also considers the coupling to an arbitrary number, $m$, of
vector multiplets, and goes on to discuss the gauged theory with an
$SU(2)$ gauge group. Various gaugings of pure $\cN=4$ supergravity
({\it i.e. not} coupled to vector or tensor multiplets) were also
considered in \LJRGgs.

The field content of the $\cN=4$ graviton multiplet is one gravitino,
four gravitinos, six vectors, four spin-$\half$ fields and a single,
real scalar. The content of the $\cN=4$ vector multiplet is: one
vector, four spin-$\half$ fields and five real scalars. The scalar
manifold of the supergravity coupled to $m$ vector multiplets is:
$$
{SO(5,m) \over SO(5) \times SO(m)} ~\times~ SO(1,1) \ ,
$$
where the factor of $SO(1,1)$ represents the scalar of the
graviton multiplet.

If one truncates the ungauged $\cN=8$ theory to the singlets of
$SU(2)_I$ one gets precisely $\cN=4$ supergravity coupled to two
$\cN=4$ vector multiplets. However the truncation of the $SO(6)$ gauged
supergravity yields something a little different from that of \MAPT\
because the residual gauge group is $SU(2)_G \times U(1)_G$ and not
just $SU(2)$. The gauging of the extra $U(1)$ appears to require the
dualization of some of the vector fields into tensor gauge fields just
as in \GRW. One can easily determine the field content and compute the
structure of the field theory by truncation the results of \GRW\ to the
$SU(2)_I$ singlet sector. Indeed, the four vectors \USPevecs\ can be
used to project the $USp(8)$ structure appropriately. To be explicit:

\item{(i)} The gravitinos of the $\cN=8$ theory
transform ${\bf 4}  \oplus {\bf \bar 4}$ of $SU(4)$.  Under
 $SU(2)_I  \times SU(2)_G \times  U(1)_G$, one has ${\bf 4} =
{\bf (2,1)}(+\half) \oplus  {\bf (1,2)}(-\half)$. Thus there are four
gravitinos that are singlets  under this $SU(2)_I$,
transforming as ${\bf 2} (-\half) \oplus{\bf 2} (+\half)$ of $SU(2)_G
\times U(1)_G$.  They are explicitly given by the
inner products of \USPevecs\ and the $\cN=8$ gravitinos.
The supersymmetries decompose similarly.

\item{(ii)} The vector
fields are in the ${\bf 15}$ of $SU(4)$ and the $SU(2)_I$ singlets are
simply the adjoints of $SU(2)_G \times U(1)_G$.

\item{(iii)}  The tensor gauge fields are in an $SL(2,\IR)$
doublet of ${\bf 6}$'s  of $SU(4)$:  and each member of the
$SL(2,\IR)$ doublet  gives rise to two tensor gauge fields that are
singlets of $SU(2)_G$ but have charges of $\pm 1$ under $U(1)_G$.

\item{(iv)}  The spin-$\half$ fields are in the following
representations of $SU(2)_G \times U(1)_G$:  ${\bf 2} (-{3 \over 2})
\oplus {\bf 2} (+{3 \over 2})  \oplus {2 \times {\bf 2} (-\half)}
\oplus 2 \times {{\bf 2} (+ \half)}$

\item{(v)}  The eleven scalars were described above.

Putting this together, the theory consists of an $\cN=4$ graviton
supermultiplet and two $\cN=4$ tensor gauge super multiplets.
The former has: one graviton, four gravitinos, four vector fields, two
tensor gauge fields, four spin-$\half$ fields, and one real scalar.
The tensor gauge multiplet has the same spin-$\half$ and scalar
content as an $\cN=4$ vector multiplet.

The truncation process can be viewed in two ways.
Mathematically one is  identifying a subset of fields whose
Lagrangian and equations of motion involve only
those fields, and such that any solution of the truncated
equations of motion can be lifted to a solution of the
untruncated equations of motion. Perhaps the simplest
way of finding such a truncation is by restricting to the
fields that are singlets under some particular symmetry
(here it is $SU(2)_I$), and then Schur's lemma guarantees
consistency with the equations of motion of the untruncated
theory.   Physically the truncated theory can be taken
as a new field theory in its own right, and it is consistent and
well-defined at least at the classical level.  Or one can regard it as 
part of
the original (or perhaps another) larger field theory.

Naively, one might hope that the $\cN=4$ truncation of gauged
$\cN=8$  supergravity corresponds to a truncation of $\cN=4$
Yang-Mills to its $SU(2)_I$ singlet sector.  However, the
$SU(2)_I$ invariant subsector of  $\cN=4$ Yang-Mills is simply
a pure $\cN=2$ Yang-Mills theory, which is not a superconformal
field theory, and so cannot, by itself be holographically dual
to the $\cN=4$  supergravity theory.  The error in this naive
hope is that fundamental fields of supergravity are not
dual to fundamental fields of Yang-Mills, and so one must
proceed more carefully.  In other words, the truncation to the $SU(2)_I$
invariant sector must be applied to the gauge invariant composite operators
rather than to the fundamental colored fields themselves.

There are however many ways to build a superconformal field theory
around the same set of $SU(2)_I$ invariant composite operators dual to
the truncated ${\cal N}=4$ supergravity theory. The focus of this paper
has been on adding adjoint hypermultiplets to make the $\cN=4$ theory,
however there also the $\cN=2$ superconformal gauge theories whose
gauge groups and matter multiplets are encoded in an ADE quiver diagram
\refs{\MDGM}. Using the $D$-brane formulation of ${\cal N}=4$
Yang-Mills theory one can make an orbifold construction
\refs{\SKES,\LNV} using a finite subgroup $\Gamma$ of $SU(2)_I$. These
finite subgroups also have an ADE classification, and indeed the ADE
orbifold leads to the corresponding ADE $\cN=2$ ``quiver theory.'' The
common subsector of all these conformal field theories is the $\Gamma$
invariant gauge-singlet operators, corresponding holographically to
{\it untwisted} closed string states. This is in a sense the ${\cal
N}=2$ Yang-Mills ``core'' of the theory. It is thus tempting to
identify this universal subsector of the conformal field theories as
the holographic dual of $\cN =4$ supergravity coupled to two tensor
multiplets. Inclusion of the twisted sectors may involve further
coupling to additional $\cN=4$ supermultiplets. The truncated ${\cal
N}=4$ supergravity is dual to the stress tensor of the full
superconformal field theory plus its superpartners under ${\cal N}=2$
superconformal invariance.

To see more explicitly how this should work, observe that the $11$
scalars of $\cN = 4$ supergravity coupled to two tensor
multiplets provides precisely the proper moduli
of the $\cN=2$ Yang-Mills vector multiplet: that is, the gauge
coupling, the $\theta$-angle, a real, symmetric $2\times 2$ mass matrix
for the two real scalars and a complex, symmetric $2\times 2$ mass
matrix for the two complex fermions. The $SU(2) \times U(1)$
$R$-symmetry can then be used to reduce the nine mass parameters to
five, and these can be represented by diagonal mass matrices with real
traces. The $U(1)$ axial symmetry can be used to reduce this by one
further parameter, leaving four independent mass perturbations, which
correspond to the four independent scalars described in \KPW.

If $\cN = 4$ supergravity coupled to  two $\cN = 4$ tensor
multiplets does indeed provide the common core of
all the quiver models, then the scalar manifold, ${\cal S}$
considered in \KPW\ is not only part of the phase diagram of
$\cN=4$ Yang-Mills, but is that part of the phase diagram that
is common to $\cN=4$ Yang-Mills and to all the $\cN=2$ quiver
theories. The critical points and flows described here would thus
also be common to all these theories.

\vfill\supereject


\appendix{A}{$SU(2)$ invariant scalar manifolds in ${\cal N}=8$
supergravity}

In this appendix we review the construction of the $SU(2)$ invariant
sectors of the scalar manifold of ${\cal N}=8$ $d=5$ gauged
supergravity which played a central role in \refs{\KPW}.  In
particular, we derive an explicit parametrization of the manifold
$\cal S$ in section~\SugraandYM\ and collect some explicit formulae
which are
used in the computation of the mass spectra in section~\sugraandopmap.

The scalar fields in ${\cal N}=8$ supergravity are described by a
non-linear $\sigma$-model on the non-compact coset space $E_{6(6)}/
USp(8)$ \refs{\ECrem}. In the gauged theory \refs{\GRWplb,\GRW,\PPvN} the
global symmetry group $E_{6(6)}$ is broken to an $SO(6)$ subgroup,
which then becomes the local gauge symmetry group of the theory.  The
authors of \refs{\KPW} considered those scalar field configurations
which are invariant under an $SU(2)$ subgroup of $SO(6)$ that we call
$SU(2)_I$. Let $C$
be the commutant of $SU(2)_I$ in $E_{6(6)}$ and $K\subset C$ the
maximal compact subgroup of $C$. Then the reduced scalar manifold
$\cal S$ corresponding to a given $SU(2)_I$ is
\eqn\scalamnf{
{\cal S}\simeq C/K \subset E_{6(6)}/USp(8)\,. }
The commutant of $SU(2)_I$ in $SO(6)$ is the group of residual
gauge transformations on $\cal S$, which will be denoted by $G$.

The scalar potential, $V$, restricted to the fields in $\cal S$, is
invariant under the residual gauge symmetry, $G$, and, in addition,
under $SL(2,\IR)$ of the ten-dimensional theory.  Because of this
invariance the potential restricted to $S$ actually depends on a
reduced number of independent parameters, denoted by ${\rm dim}\,V$
which is less than ${\rm dim}\,S$.

There are four inequivalent embeddings of $SU(2)_I\hookrightarrow
SO(6)$ corresponding to the following branchings of the fundamental
and the vector representations of $SO(6)$:
$$
\seqalign{
\span\TC\qquad &  \span\TR\qquad &  \span\TR}
{{\rm (i)} & \hbox{
$\bf 4\rightarrow  2\oplus 1\oplus 1$}\,, &
        \hbox{
$\bf 6\rightarrow  2\oplus 2\oplus 1\oplus 1$}\,;\cr
{\rm (ii)} & \hbox{
        $\bf 4\rightarrow 2\oplus 2$}\,,& \hbox{
        $\bf 6\rightarrow 3\oplus 1\oplus 1\oplus 1$}\,;\cr
{\rm (iii)} & \hbox{
        $\bf 4 \rightarrow 3\oplus 1$}\,, &\hbox{
        $\bf 6\rightarrow  3\oplus 3$}\,;\cr
{\rm (iv)} & \hbox{
        $\bf 4\rightarrow 4$}\,, &\hbox{
        $\bf 6\rightarrow  5\oplus 1$}\,.
}
$$
The resulting scalar manifolds, $\cal S$, are summarized in Table~A.1.
One should note  that this table gives the local
geometry
of $\cal S$, but does not capture discrete symmetries that might be
present.

Given an explicit realization of $E_{6(6)}$, it is rather
straightforward to derive a parametrization of $\cal S$ once the
subgroup $SU(2)_I$ has been identified. We will illustrate this for
case (i), which is the relevant one for this paper.  Other cases can
be worked out similarly.

\goodbreak\midinsert
$$
\seqalign{
\span\TC\quad &  \span\TC\qquad &  \span\TC\qquad &  \span\TC\qquad
&  \span\TC\quad &  \span\TC }
{SU(2)_I & C & K & G & {\rm dim}\,{\cal S} & {\rm dim}\, V \cr
\noalign{\medskip\hrule\medskip}
{\rm (i)} & SO(5,2)\times O(1,1) & SO(5)\times SO(2) & SU(2) \times U(1) &
11
& 4 \cr
{\rm (ii)} & SL(3,\IR)\times SL(3,\IR)  & SO(3)\times SO(3)
& SU(2) & 10 & 4 \cr
{\rm (iii)} & G_{2(2)} & SO(3)\times SO(3) & U(1)\times U(1) & 8 & 4
\cr
{\rm (iv)} & O(1,1) \times SL(2,\IR) & - & - & 4 & 1 \cr
\noalign{\medskip\hrule}
}
$$
\vskip -0.5cm
\noindent{\ninepoint\sl \baselineskip=8pt {\bf Table A.1}:
Scalar manifolds, $\cal S$, for inequivalent  embeddings
$SU(2)_I\hookrightarrow  SO(6)$.}
\medskip
\endinsert

Our starting point is a real 27-dimensional realizations of $E_{6(6)}$
in the $SL(6,\IR)\times SL(2,\IR)$ basis%
\foot{We use  the notation $I,J,\ldots =1,\ldots, 6$ and
$\al,\be,\ldots
=1,2$ for the $SL(6,\IR)$ and $SL(2,\IR)$ indices, respectively.}
$(z_{IJ},z^{I\al})$, $z_{IJ}=-z_{JI}$, as described in \refs{\ECrem}
and, in particular, in Appendix~A of \refs{\GRW}, which the reader
should consult for conventions and further details.  An infinitesimal
$E_{6(6)}$ action on $(z_{IJ},z^{I\al})$ is given by
\eqn\esixtran{\eqalign{
\delta z_{IJ}&=-\La^K{}_Iz_{KJ}-\La^K{}_Jz_{IK}+\Si_{IJK\be}
z^{K\be}\,,\cr
\delta z^{I\al}&=\La^I{}_Kz^{K\al}+\La^\al{}_\be z^{I\be}+\Si^{KLI\al}
z_{KL}\,,\cr}
}
where $\La^I{}_J$, $\La^\al{}_\be$ and $\Si_{IJK\al}={1\over 6}
\epsilon_{IJKLMN} \epsilon_{\al\be}\Si^{LMN\be}$ correspond to
$SL(6,\IR)$, $SL(2,\IR)$ and the coset elements, respectively. The
antisymmetric matrices $\La^I{}_J=-\La^J{}_I$ generate the gauge group
$SO(6)$.

Define $4\times 4$ 't Hooft matrices
\eqnn\THmatr
$$
\eta_1^{(\pm)}\eql{1\over 2} \left(\matrix{&1&&\cr -1&&&\cr &&&\pm 1\cr
&&\mp 1&\cr}\right)\,, \qquad
\eta_2^{(\pm)}\eql{1\over 2} \left(\matrix{&&\mp 1&\cr &&&1\cr \pm
1&&&\cr
&-1&&\cr}\right)
\,,\eqno{\THmatr} $$
$$
\eta_3^{(\pm)}\eql{1\over 2} \left(\matrix{&&&1\cr &&\pm 1&\cr &\mp
1&&\cr
-1&&&\cr}\right)
\,,$$
satisfying
\eqn\THcomm{
[\eta^{(+)}_r,\eta^{(-)}_s]\eql 0\,,\qquad
[\eta^{(\pm)}_r,\eta^{(\pm)}_s]\eql \epsilon_{rst} \eta_t^{(\pm)}\,,
}
and
\eqn\THselfdual{
{1\over 2} \epsilon_{ijkl}\eta^{(\pm)}_r{}^{kl} \eql
\pm\eta^{(\pm)}_r{}^{ij}\,.
}
The $SU(2)_I$ in case (i) is generated by the $SO(6)$ matrices
\eqn\TheTHemb{
(\La^I{}_J)=\left(\matrix{\eta_r^{(-)} & 0 & 0 \cr
0& 0 & 0 \cr 0& 0 & 0 \cr}\right)\,,\qquad r=1,2,3\,.
}
The residual gauge group is $SU(2)_G\times U(1)_G$, where $SU(2)_G$ is
generated by the $\eta_r^{(+)}$ in the same $4\times 4$ block as
above, so that $SU(2)_I\times SU(2)_G$ is the obvious $SO(4)$ subgroup
of $SO(6)$, while $U(1)_G$ are the $SO(2)$ rotations in the 56-block.

Now, let us introduce a new basis in which the $SO(5,2)$
transformations acquire a canonical form.  A helpful observation here
is the branching of the $\bf 27$ of $E_{6(6)}$ under $SU(2)_I\times
SO(5,2)\times O(1,1)$,
\eqn\theesbr{
{\bf  27}\quad \rightarrow\quad
{\bf (1,1)}(-4)\oplus
{\bf (3,1)}(2)\oplus
{\bf (1,7)}(2)\oplus
{\bf (2,8)}(-1)\,.
}
The corresponding new basis  is explicitly given by
\eqnn\nbassing \eqnn\nbastrip \eqnn\nbasvect
\eqnn\nbasspone \eqnn\nbassptwo
$$\eqalignno{
u_{(1,1)}& \eql z_{56}\,;&\nbassing \cr
\noalign{\smallskip}
u_{(3,1)}^1& \eql{1\over \sqrt{2}}(z_{12}-z_{34})\,,\quad
u_{(3,1)}^2\eql{1\over \sqrt{2}}(z_{13}+z_{24})\,,\quad
u_{(3,1)}^3\eql{1\over \sqrt{2}}(z_{14}-z_{23})\,;& \nbastrip\cr
u_{(1,7)}^1& \eql-{1\over \sqrt{2}}(z_{12}+z_{34})\,,\quad
u_{(1,7)}^2\eql{1\over \sqrt{2}}(z_{13}-z_{24})\,,\quad
u_{(1,7)}^3\eql-{1\over \sqrt{2}}(z_{14}+z_{23})\,,& \cr
u_{(1,7)}^4& \eql{1\over 2}(z^{51}-z^{62})\,,\quad
u_{(1,7)}^5\eql{1\over 2}(z^{52}+z^{61})\,,\quad
u_{(1,7)}^6\eql{1\over 2}(z^{51}+z^{62})\,,& \nbasvect \cr
u_{(1,7)}^7& \eql-{1\over 2}(z^{52}-z^{61})\,;& \cr
u_{(2,8)}^{11}&\eql {1\over \sqrt{2}}(z_{15}-i z_{25})\,,\quad
u_{(2,8)}^{12}\eql {1\over \sqrt{2}}(z_{16}-i z_{26})\,,\quad
u_{(2,8)}^{13}\eql {1\over \sqrt{2}}(z_{35}+i z_{45})\,,&\cr
u_{(2,8)}^{14}&\eql {1\over \sqrt{2}}(z_{36}+i z_{46})\,,\quad
u_{(2,8)}^{15}\eql {1\over 2}(z^{21}+i z^{11})\,,\quad
u_{(2,8)}^{16}\eql {1\over 2}(z^{22}+i z^{12})\,,& \nbasspone \cr
u_{(2,8)}^{17}&\eql {1\over 2}(z^{41}-i z^{31})\,,\quad
u_{(2,8)}^{18}\eql {1\over 2}(z^{42}-i z^{32})\,;& \cr
u_{(2,8)}^{21}&\eql -{1\over \sqrt{2}}(z_{35}-i z_{45})\,,\quad
u_{(2,8)}^{22}\eql -{1\over \sqrt{2}}(z_{36}-i z_{46})\,,\quad
u_{(2,8)}^{23}\eql {1\over \sqrt{2}}(z_{15}+i z_{25})\,,&\cr
u_{(2,8)}^{24}&\eql {1\over \sqrt{2}}(z_{16}+i z_{26})\,,\quad
u_{(2,8)}^{25}\eql -{1\over 2}(z^{41}+i z^{31})\,,\quad
u_{(2,8)}^{26}\eql -{1\over 2}(z^{42}+i z^{32})\,,&\nbassptwo\cr
u_{(2,8)}^{27}&\eql {1\over 2}(z^{21}-i z^{11})\,,\quad
u_{(2,8)}^{28}\eql {1\over 2}(z^{22}-i z^{12})\,.&\cr
}
$$
We will denote the elements of this basis by $u^\Omega$,
$\Omega=1,\ldots, 27$ and define the transition matrix $O$,
\eqn\thematO{z_{IJ}\eql O_{IJ\,\Omega} u^\Omega\,,\qquad
z^{I\alpha}\eql
O^{I\alpha}{} _\Omega u^\Omega\,,}
and its inverse $\widetilde O$,
\eqn\themattO{
u^\Omega\eql {1\over 2} \tO^{\Omega\,IJ}
        z_{IJ}+\tO^{\Omega}{}_{I\alpha}z^{I\alpha}\,.}

 The eleven-dimensional family of infinitesimal noncompact transformations
of the group $SO(5,2)\times O(1,1)$ is parametrized by $x^{i},\,y^i$,
$i=1,\ldots,5$ and $\al$, such that
 \eqn\thecommso{
(\La^I{}_J)\eql {\rm diag}(-\alpha,-\alpha,-\alpha,-\alpha,
\left(\matrix{2\alpha-x^4+y^5&-x^5-y^4\cr -x^5-y^4 &
2\alpha+x^4-y^5\cr}\right))\,,
}
\eqn\thecomtwo{
(\La^\alpha{}_\beta)\eql \left(\matrix{-x^4-y^5& -x^5+y^4\cr
         -x^5+y^4& x^4+y^5\cr}\right)\,,
}
and
\eqn\thecommth{
\Sigma^{IJ51}\eql\Sigma^{IJ62}\eql  \sum_{i=1}^3 \sqrt{2}
x^i\eta_i^{IJ}\,,\quad \Sigma^{IJ61}\eql-\Sigma^{IJ52}\eql
\sum_{i=1}^3 \sqrt{2}
y^i\eta_i^{IJ}\,,\quad I,J=1,\ldots,4\,,} where $\eta_r=\eta_r^{(+)}$.
It is now easy to verify that these transformations commute with
$SU(2)_I$. Moreover, we find that in the new basis they become
\eqn\trffone{
\delta u_{(1,1)}\eql -4\alpha u_{(1,1)}\,,
}
\eqn\trfftwo{
\delta u_{(3,1)}^\rho\eql 2\alpha u_{(3,1)}^\rho\,,
}
\eqn\trffthree{
\delta u_{(1,7)}^i\eql 2\alpha u_{(1,7)}^i+M_7(x,y)^i{}_j
u_{(1,7)}^j\,,
}
and
\eqn\trfffour{
\delta u_{(2,8)}^{1,2}\eql -\alpha u_{(2,8)}^{1,2}+M_8(x,y)
u_{(2,8)}^{1,2}\,,
}
where
\eqn\trfffive{
M_8(x,y)\eql \sum_{i=1}^5
(x^i\ggamma_i\ggamma_6+y^i\ggamma_i\ggamma_7)\,,
}
and $M_7(x,y)$ is the corresponding vector matrix
\eqn\trffsix{
M_7(x,y)\eql-2 \left(\matrix{0&x&y\cr
x^T&0&0\cr
y^T&0&0\cr}\right)
}
satisfying
\eqn\trffseven{
M_8(x,y)\ggamma_i-\ggamma_iM_8(x,y)\eql M_7(x,y)^j{}_i\ggamma_j\,.  }
The $SO(5,2)$ gamma matrices $\ggamma_i$, $i=1,\ldots,7$, are given in
Appendix C. Note that $u_{(2,8)}^1$ and $u_{(2,8)}^2$ form a pair of
symplectic Majorana spinors, $u_{(2,8)}^2=i\ggamma_2 (u_{(2,8)}^1)^*$,
as required by the reality of the $\bf 27$ of $E_{6(6)}$.

Let us write \trffone-\trfffour\ in the matrix form $\delta
u^\Om=M(x,y,\al)^\Om{}_\Si U^\Si$. The reduced scalar manifold $\cal S$
consists of  finite transformations
\eqn\thefinD{
D(x,y,\al)=\exp M(x,y,\al)\,,
}
and by the standard properties of the non-compact cosets \refs{\Helg}
has global coordinates  $x^i$, $y^i$ and $\al$.

To define the $27$-bein on $\cal S$ we  introduce yet another
basis, $z^{ab}$, $a,b=1,\ldots,8$, transforming in the $\bf 27$ of
$USp(8)$. It is defined in terms of the $SL(6,\IR)\times SL(2,\IR)$ basis
by (see, (A.45) of \refs{\GRW})
\eqn\uspbasis{
z^{ab}\eql {1\over 4}(\Ga_{IJ})^{ab}z_{IJ}+{1\over 2\sqrt{2}}
(\Ga_{I\al})^{ab}z^{I\al}\,,} where the gamma matrices $\Gamma_{IJ}$
and $\Gamma_{I\alpha}$ are given in Appendix C.1.  By combining
transformations between the various bases, we find that the $27$-bein
$(\calV^{IJ\,ab},\calV_{I\al}{}^{ab})$ on $\cal S$ is
\eqn\appcviel{
\calV^{IJ\,ab}\eql {1\over 2}
        G^{ab}{}_\Om D^{\Omega}{}_\Sigma\, \tO^{\Si\,IJ}\,,
\qquad
\calV_{I\al}{}^{ab}\eql G^{ab}{}_\Om
D^\Om{}_\Si\,\tO^\Si{}_{I\al}\,,}
where
\eqn\transss{
G^{ab}{}_\Om\eql
{1\over 4} (\Gamma_{IJ})^{ab}O_{IJ\,}{}_\Omega +
{1\over 2\sqrt{2}}(\Gamma_{I\al})^{ab} O^{I\al}{}_\Omega\,,
}
while the inverse  27-bein is
\eqn\inverviel{
\widetilde \calV_{ab\,IJ}\eql
        O_{IJ\,\Om}\widetilde D^\Om{}_\Si \widetilde G^\Si{}_{ab}\,,
\qquad
\widetilde \calV_{ab}{}^{I\al}\eql
O^{I\al}{}_\Om \widetilde D^\Om{}_\Si \widetilde G^\Si{}_{ab}\,,
}
with
\eqn\invtrans{
\widetilde G^\Om{}_{ab}\eql {1\over 8} \widetilde O^{\Om\,IJ}
(\Ga_{IJ})^{ab}-{1\over 2\sqrt{2}}\widetilde O^\Om{}_{I\al}
(\Ga_{I\al})^{ab}\,.
}

The potential, $V$, of gauged ${\cal N}=8$ is defined in terms of
$USp(8)$ tensors
\eqn\theWten{
W_{abcd}=\epsilon^{\al\be}\delta^{IJ} \calV_{I\al ab}\calV_{J\be cd}\,,
\qquad W_{ab}=W^c{}_{acb}\,,}
where the $USp(8)$ indices are lowered using the symplectic
metric $\Omega_{ab}$. We have  \refs{\GRW}
\eqn\Potential{
V=-\hbox{$1\over 32$}g^2 \left(2 (W_{ab})^2-(W_{abcd})^2\right)\,.
}

In general, the potential, $V$, is invariant under the $G_R\times G_L$
group of local transformations of the 27-bein, where $G_R=USp(8)$ and
$G_L=SO(6)\times SL(2,\IR)$. Upon restriction to the scalar submanifold,
$\cal S$, a  subgroup of those transformations
\eqn\theresgg{
D\quad \rightarrow \quad G'DG'{}^{-1} P\,,} preserves $\cal S$ and
leaves $V$ invariant. The transformations in $G'$ act linearly on
$x^i$ and $y^i$. In terms of the $5\times 2$ matrix $(x,y)$ in
\trffsix\ this action is schematically  of the form
\eqn\thegpact{
(x,y)\quad \rightarrow \quad \left( \matrix{ R_{SO(3)} & \cr &
R_{SO(2)} \cr}\right) (x,y) (R_{SO(2)'})\,, }
where the $SO(3)$ subgroup are the $SO(5,2)$ rotations in the
$(123)$-hyperplane, the $SO(2)$ are the rotations in the (45)-plane
and $SO(2)'$, which rotates $(x^i,y^i)$ as a doublet, are the
rotations in the (67)-plane. The group of residual gauge
transformations, $G$, in Table A.1 consists of $SO(3)$ and a diagonal
SO(2) in $SO(2)\times SO(2)'$ above. Another combination of the
$SO(2)$ rotations is a part of $SL(2,\IR)$. An infinitesimal action of
$P$,
the noncomapct generators of $SL(2,\IR)$, with parameters $\varepsilon^1$
and $\varepsilon^2$ on $(x,y)$ is given by
\eqn\noncomact{
\left(\matrix{x^4 & y^4\cr x^5 & y^5\cr}\right)
\quad \rightarrow \quad
\left(\matrix{x^4+\varepsilon^1 & y^4+ \varepsilon^2\cr
x^5-\varepsilon^2 & y^5+\varepsilon^1\cr}\right)\,.}
It was argued in \refs{\KPW} that using those invariances one may set
all parameters, but $x^1$, $x^4$, $y^2$ and $\al$, to zero, thus
reducing the dependence of the potential from 11 to 4 parameters. This
simplification allowed the authors of \refs{\KPW} to compute the
potential explicitly in terms of the variables $r_x$, $r_y$, $\theta$
and $\alpha$, where
\eqn\newvars{
x^1=r_x\cos\theta\,,\quad x^4=r_x\sin\theta\,,\quad y^2=r_y\,.
}

The ${\cal N}=2$ critical point is at $r_x=r_y=\pm{1\over 4}\log(3)$,
$\theta=0$ and $\alpha=\alpha_0={1\over 6}\log (2)$. It is invariant
under $SU(2)_I\times U(1)_R$ subgroup of $SO(6)$, where the $R$-charge
subgroup $U(1)_R$ is an unbroken subgroup of $SU(2)_G\times U(1)_G$.
To identify this subgroup we note that the matrix $M(x,y,\alpha)$ at
the critical point, when expressed in the $SL(6,\IR)\times SL(2,\IR)$
basis using \uspbasis-\transss, is given by
\eqn\thecritva{
 (\La^I{}_J)\eql {\rm
diag}(-\al_0,-\al_0,-\al_0,-\al_0,2\al_0,2\al_0)\,,\qquad
 (\La^\al{}_\be) \eql
(0)\,, }
and
\eqn\thecritvb{
\Si^{IJ51}\eql \Si^{IJ62}\eql \pm {1\over 2\sqrt{2}}\log(3)
\eta^{(+)}_1{}^{IJ}\,,\qquad
\Si^{IJ61}\eql- \Si^{IJ52}\eql \pm {1\over 2\sqrt{2}}\log(3)
\eta^{(+)}_2{}^{IJ}\,.
}
The only nontrivial condition is that the $\Si$'s above are
invariant under $U(1)_R$, which determines the $R$-charge generator
\eqn\Rcharge{
Y_R\eql (\La^I{}_J)\eql \left(
\matrix{ \eta_3^{(+)} & 0 & 0 \cr
        0 & 0 & -1 \cr 0 & 1 & 0 \cr}\right)\,.  } It has been
normalized such that under $SU(2)_I\times U(1)_R\hookrightarrow SO(6)$
we have
\eqn\thesixbranch{
{\bf 6}\quad\rightarrow\quad {\bf 2}_{1/2}\oplus {\bf 2}_{-1/2}\oplus
{\bf 1}_1 \oplus {\bf 1}_{-1}\,.}

Finally, we compute the $USp(8)$ tensors $W_{ab}$, $W_{abcd}$ and
\eqn\theAtensor{
A_{abcd}=-3W_{a[bcd]|}\,,
}
at the critical point.
In particular,
\eqn\theWtwoab{
(W_{ab})= {1\over
6 \cdot 2^{1/3}} \left( \matrix{-15& & & & & -1& & \cr & -15& & & 1& & & \cr
& & -13& & & & & -1\cr & & & -13& & & 1& \cr & 1& & & -15& & & \cr -1&
& & & & -15& & \cr & & & 1& & & -13& \cr & & -1& & & & &
-13\cr}\right) \,.
}

\appendix{B}{Unitary multiplets of $SU(2,2|1)$}

In this appendix we recall the classification of the unitary highest
weight representations of the superalgebra $SU(2,2|1)$ in
\refs{\FlatFrons,\DobrPetk} (see, also \refs{\KacWak}) and work out explicitly
shortening patterns for the multiplets that are relevant for the
discussion in section~\sugraandopmap.

A unitary highest weight representation of $SU(2,2|1)$ can be
decomposed into a direct sum of unitary highest weight representations
of its bosonic subalgebra $SU(2,2)\times U(1)$.  The latter has a
maximal compact subalgebra $U(1)\times SU(2)\times SU(2)\times U(1)$,
in which the generator of the first $U(1)$ factor is the energy and
that of the second is $R$-charge. The two $SU(2)$ factors give the
obvious $SO(4)$ subalgebra of $SU(2,2) \simeq SO(4,2)$.  Highest
weight representations, ${\cal D}(E_0,j_1,j_2;r)$, of $SU(2,2|1)$ are
parametrized by the corresponding four quantum numbers.

\proclaim Theorem \refs{\FlatFrons,\DobrPetk}. {A representation,
${\cal D}(E_0,j_1,j_2;r)$, is unitary if and only if one of the
following conditions (i)-(iv) holds:\smallskip
\item{(i)} $E_0\geq 2j_1-{3\over 2}r+2$ and
        $E_0\geq 2j_2+{3\over 2}r+2$ for $j_1,j_2\geq0$,
\item{(ii)} $E_0=-{3\over 2}r$ and
$E_0\geq 2j_2+{3\over 2}r+2$ for $j_1=0$ and $j_2\geq 0$,
\item{(iii)} $E_0={3\over 2}r$ and
$E_0\geq 2j_1-{3\over 2}r+2$ for $j_1\geq 0$ and $j_2=0$,
\item{(iv)} $E_0=r=0$ for $j_1=j_2=0$.
}
\smallskip

Maximal representations (long multiplets) consist of the 16
representations of $SU(2,2)$ listed in the second column of 
Table~B.1, each with multiplicity 1. One may notice that each
representation in the multiplet is uniquely identified by its
$(j_1,j_2;r)$. The level in the first column of the table is the
minimal number of supersymmetry operators that must be applied to the
vacuum to reach a given representation.

\goodbreak\midinsert
$$
\seqalign{
\span\TC\qquad &  \span\TT\qquad\qquad & \span\TC}
{{\rm Level} & $SU(2,2)$ representation & {\rm Multiplicity} \cr
\noalign{\medskip\hrule\medskip}
0 & $D(E_0,j_1,j_2;r)$  & 1\cr
\noalign{\medskip\hrule\medskip}
1 & $D(E_0+\half,j_1+\half,j_2;r-1)$ & \nn1+\cr
& $D(E_0+\half,j_1-\half,j_2;r-1)$ & s_1 \nn1-\cr
& $D(E_0+\half,j_1,j_2-\half;r+1)$ & s_2 \nn2-\cr
& $D(E_0+\half,j_1,j_2+\half;r+1)$ & \nn2+\cr
\noalign{\medskip\hrule\medskip}
2 & $D(E_0+1,j_1,j_2;r-2)$ & \nn1+\nn2-\cr
  & $D(E_0+1,j_1+\half,j_2+\half;r)$ & \nn1+\nn2+\cr
  & $D(E_0+1,j_1+\half,j_2-\half;r)$ & s_2\nn1+\nn2-\cr
  & $D(E_0+1,j_1-\half,j_2+\half;r)$ & s_1\nn1-\nn2+\cr
  & $D(E_0+1,j_1-\half,j_2-\half;r)$ & s_1s_2\nn1-\nn2-\cr
  & $D(E_0+1,j_1,j_2;r+2)$ & \nn1-\nn2+\cr
\noalign{\medskip\hrule\medskip}
3 & $D(E_0+{3\over 2},j_1,j_2+\half;r-1)$ & \nn1+\nn1-\nn2+\cr
  & $D(E_0+{3\over 2},j_1,j_2-\half;r-1)$ & s_1 \nn1-\nn2+\nn2-\cr
  & $D(E_0+{3\over 2},j_1-\half,j_2;,r+1)$ & s_2 \nn1+\nn1-\nn2-\cr
  & $D(E_0+{3\over 2},j_1+\half,j_2;r+1)$ & \nn1+\nn2+\nn2-\cr
\noalign{\medskip\hrule\medskip}
4 & $D(E_0+{ 2},j_1,j_2;r)$ & \nn1+\nn1-\nn2+\nn2-\cr
\noalign{\medskip\hrule}
&&
}
$$
\leftskip 2pc\rightskip 2pc
\noindent{\ninepoint\sl \baselineskip=8pt {\bf Table B.1}:
Multiplicities of $SU(2,2)$ representations in the decomposition of
an $SU(2,2|1)$ unitary representation ${\cal D}(E_0,j_1,j_2;r)$.}
\endinsert
\medskip

Shorter multiplets arise when either $j_1j_2=0$, or, more
interestingly, when some of the equalities in the unitarity bounds
(i)-(iii) are saturated. In the latter case some of the descendant
$SU(2,2)$ representations that are predicted by the tensor product
rules are missing. A complete list of shortening patterns is given in
\refs{\FlatFrons}. One can succinctly summarize the structure of the
resulting short modules as follows:

\proclaim Corollary.
 Define $N(x)=1$ for $x\not=0$ and $N(0)=0$. Let
\eqn\varens{\eqalign{
\nn1+=N(E_0+2j_1+{3\over 2}r)\,,\qquad &\nn1-=N(E_0-2j_1+{3\over
2}r-2)\,,\cr
\nn2+=N(E_0+2j_2-{3\over 2}r)\,,\qquad &\nn2-=N(E_0-2j_2-{3\over
2}r-2)\,,\cr}
}
and
\eqn\spinvan{
s_1=N(j_1)\,,\qquad s_2=N(j_2)\,. } The multiplicities of the
$SU(2,2)$ representations in the decomposition of a unitary
representation, ${\cal D}(E_0,j_1,j_2;r)$, are determined by \varens,
\spinvan\ and the appropriate products of these numbers as given  in
Table B.1.
\smallskip

The short multiplets in Table~6.1 in section~\sugraandopmap\
 are either chiral
(complex) or non-chiral (real). The shortening conditions for the
chiral multiplets are:\smallskip
\item{(R)}  $E_0=-{3\over 2}r$ and $j_1=0$ for the RH-multiplets,
\item{(L)}  $E_0={3\over 2}r$ and $j_2=0$ for the LH-multiplets.
\smallskip
\noindent
We then have $\nn1+=0$ and $\nn2+=0$, respectively.  For a generic
value of $E_0$, a LH-multiplet is a direct sum of four $SU(2,2)$
multiplets for $j>0$ and three multiplets for $j=0$ as shown in
Table~B.2. The RH-multiplets are obtained by $(j_1,j_2)\rightarrow
(j_2,j_1)$ and $r\rightarrow -r$.

\goodbreak\midinsert
\medskip
$$
\seqalign{
\span\TC\qquad\qquad  &  \span\TC\qquad &  \span\TC\qquad &
\span\TC }
{E\,\backslash \,R & r & r-1 & r-2   \cr
\noalign{\medskip\hrule\medskip}
E_0 & (j,0) & & \cr
E_0+\half & &(j+\half,0)\oplus (j-\half,0) & \cr
E_0+1 & & & (j,0)\cr
\noalign{\medskip\hrule\medskip}
}
$$
\leftskip 2pc\rightskip 2pc
\vskip -0.5cm
\noindent{\ninepoint\sl \baselineskip=8pt {\bf Table B.2}:
Chiral LH-multiplets ${\cal D}(E_0,j,0;r)$, where $r={2\over
3}E_0$.  }
\medskip
\endinsert

The non-chiral short multiplets are obtained by setting $j_1=j_2=j$,
$E_0=2j+1$ and $r=0$ so that $\nn1-=\nn2-=0$.  The multiplet is shown
in Table B.3. For a complete list of massless chiral and non-chiral
short multiplets with spins not exceeding 2, we refer the reader to,
e.g., Table~5 in \refs\GRW.

\goodbreak\midinsert
\medskip
$$
\seqalign{
\span\TC\qquad\qquad  &  \span\TC\qquad &  \span\TC\qquad &
\span\TC }
{E\,\backslash\,R & -1 & 0 & 1   \cr
\noalign{\medskip\hrule\medskip}
E_0 &  &(j,j) & \cr
E_0+\half & (j+\half,j) &  & (j,j+\half) \cr
E_0+1 & & (j+\half,j+\half) & \cr
\noalign{\medskip\hrule\medskip}}
$$
\leftskip 2pc\rightskip 2pc
\vskip -0.5cm
\noindent{\ninepoint\sl \baselineskip=8pt {\bf Table B.3}:
Non-chiral multiplets ${\cal D}(E_0,j,j;0)$, where $E_0=2j+2$.  }
\medskip
\endinsert

The multiplet with irrational $E_0$ found in section~\sugraandopmap\
 is an
example of a long representation with only nine $SU(2,2)$
representations because its vacuum representation has $j_1=j_2=0$.

Finally, there are multiplets with the simplest type of shortening
when only one vanishing condition holds, so that the resulting
multiplet extends through four energy levels, rather than three as in
the case of the usual short multiplets. The defining vanishing
relation for the modules of this type which arise in section
\sugraandopmap\
 are:
$E_0=2j_1-{3\over 2}r+2$ or $E_0=2j_2+{3\over 2}r+2$, which is the same
as
$\nn1-=0$ or $\nn2-=0$.  We will refer to them as RH
and LH semi-long multiplets.  A typical LH-semi-long multiplet is of
the form given in Table B.4.

\goodbreak\midinsert
\medskip
$$
\seqalign{
\span\TC\qquad\qquad  &  \span\TC\qquad &  \span\TC\qquad
&  \span\TC\qquad &  \span\TC }
{E\,\backslash\,R & r+1 & r & r-1  & r-2  \cr
\noalign{\medskip\hrule\medskip}
E_0 &  &  (j_1,j_2) &  & \cr
E_0+\half & (j_1,j_2+\half) &   & (j_1+\half,j_2) &  \cr
          & &  & (j_1-\half,j_2) & \cr
E_0+1 &  & (j_1+\half,j_2+\half) & & (j_1,j_2) \cr
 & &(j_1-\half,j_2+\half) &  & \cr
E_0+\hbox{${3\over 2}$} & & & (j_1,j_2+\half) & \cr
\noalign{\medskip\hrule}
}
$$
\leftskip 2pc\rightskip 2pc
\vskip -0.5cm
\noindent{\ninepoint\sl \baselineskip=8pt {\bf Table B.4}:
LH-semi-long multiplet ${\cal D}(E_0,j_1,j_2;r)$, where
$r={2\over 3}(E_0-2j_2-2)$.  }
\medskip
\endinsert

\noindent
The corresponding RH-semi-long multiplet is obtained from the above
by $(j_1,j_2)\rightarrow (j_2,j_1)$ and $r\rightarrow -r$. Once more
there might be a further shortening of semi-short multiplets for
$j_1$ or $j_2$ equal zero. An example is the last multiplet in
Table~6.2.

\appendix{C}{$SO(7)$ and $SO(5,2)$ gamma matrices}

In this appendix we summarize explicit realizations of the $SO(7)$ and
$SO(5,2)$ gamma matrices, which are needed to derive some of the
formulae in section~\sugraandopmap\ and Appendix~A.
\medskip
\noindent
\line{\it C.1. $SO(7)$ gamma matrices\hfill }
\medskip
We follow here the same convention as in Appendix~A of \refs{\GRW},
where the $SO(7)$ gamma matrices, $\Gamma_i=(\Gamma_i{}^{ab})$,
$i=0,1,\ldots,6$, are
hermitian and  skew symmetric (\ie\ pure imaginary) and satisfy
\eqn\appcGa{
\{\Gamma_i,\Gamma_j\}=\delta_{ij}\,,
}
\eqn\appcGapr{
\Gamma_0=i\,\Gamma_1\Gamma_2\Gamma_3\Gamma_4\Gamma_5\Gamma_5\,.
}
A particular realization with these properties is given by
$$
\Gamma_{1}=\left(\matrix{& -i& & & & & & \cr i& & & & & & & \cr
  & & & -i& & & & \cr & & i& & & & & \cr
  & & & & & i& & \cr & & & & -i& & & \cr
  & & & & & & & i\cr & & & & & & -i& \cr }\right), \quad
\Gamma_{2}=\left(\matrix{& & i& & & & & \cr & & & -i& & & & \cr
  -i& & & & & & & \cr & i& & & & & & \cr
  & & & & & & -i& \cr & & & & & & & i\cr
  & & & & i& & & \cr & & & & & -i& & \cr }\right),
$$
$$
\Gamma_{3}=\left(\matrix{& & & -i& & & & \cr & & -i& & & & & \cr
  & i& & & & & & \cr i& & & & & & & \cr
  & & & & & & & i\cr & & & & & & i& \cr
  & & & & & -i& & \cr & & & & -i& & & \cr }\right) ,\quad
\Gamma_{4}=\left(\matrix{& & & & & -i& & \cr & & & & i& & & \cr
  & & & & & & & i\cr & & & & & & -i& \cr
  & -i& & & & & & \cr i& & & & & & & \cr
  & & & i& & & & \cr & & -i& & & & & \cr }\right) ,
$$
$$
\Gamma_{5}=\left(\matrix{& & & & & & -i& \cr & & & & & & & -i\cr
  & & & & i& & & \cr & & & & & i& & \cr
  & & -i& & & & & \cr & & & -i& & & & \cr
  i& & & & & & & \cr & i& & & & & & \cr }\right),\quad
\Gamma_{6}=\left(\matrix{& & & & & & & -i\cr & & & & & & i& \cr
  & & & & & -i& & \cr & & & & i& & & \cr
  & & & -i& & & & \cr & & i& & & & & \cr
  & -i& & & & & & \cr i& & & & & & & \cr }\right) .
$$

We also define
\eqn\appccomm{
\Gamma_{IJ}={1\over 2}(\Gamma_I\Gamma_J-\Gamma_J\Gamma_I)\,,
}
and
\eqn\appccomz{
\Gamma_{I\alpha}=(\Gamma_I,i\Gamma_I\Gamma_0)\,,
}
for $I,J=1,\ldots,6$ and $\alpha=1,2$.

The $USp(8)$ indices, $a,b,\ldots=1,\ldots,8$, are raised and lowered
using the symplectic form
\eqn\appcsymform{
\Omega_{ab}=-\Omega^{ab}=i\Gamma_0{}^{ab}\,,
}
which is given by
$$
(\Omega_{ab})=
\left(\matrix{
        & & & & 1& & & \cr & & & & & 1& & \cr
  & & & & & & 1& \cr & & & & & & & 1\cr
  -1& & & & & & & \cr & -1& & & & & & \cr
  & & -1& & & & & \cr & & & -1&\phantom{-1} &\phantom{-1} &\phantom{-1}
 &\phantom{-1} \cr}\right)\,.
$$
We refer the reader to \refs{\GRW} for further details.
\bigskip
\noindent
\line{\it C.2. $SO(5,2)$ gamma matrices\hfill }
\medskip
The $SO(5,2)$ gamma matrices, $\widetilde \Gamma_i$, $i=1,\ldots,7$,
satisfy
\eqn\appcsofm{
\{\widetilde \Gamma_i,\widetilde \Gamma_j\}=2\eta_{ij}\,,
}
where $(\eta_{ij})={\rm diag}(1,1,1,1,1,-1,-1)$. We use the following
representation:

$$
\widetilde\Gamma_1=\left(\matrix{
1&&&&&&&\cr
&1&&&&&&\cr
&&-1&&&&&\cr
&&&-1&&&&\cr
&&&&-1&&&\cr
&&&&&-1&&\cr
&&&&&&1&\cr
&&&&&&&1\cr }\right),\quad
\widetilde\Gamma_2=\left(\matrix{
&&i&&&&&\cr
&&&i&&&&\cr
-i&&&&&&&\cr
&-i&&&&&&\cr
&&&&&&i&\cr
&&&&&&&i\cr
&&&&-i&&&\cr
&&&&&-i&&\cr }\right),
$$

$$
\widetilde\Gamma_3=(-1)\left(\matrix{
&&1&&&&&\cr
&&&1&&&&\cr
1&&&&&&&\cr
&1&&&&&&\cr
&&&&&&1&\cr
&&&&&&&1\cr
&&&&1&&&\cr
&&&&&1&&\cr }\right),\quad
\widetilde\Gamma_4=\left(\matrix{
&&&&1&&&\cr
&&&&&-1&&\cr
&&&&&&-1&\cr
&&&&&&&1\cr
1&&&&&&&\cr
&-1&&&&&&\cr
&&-1&&&&&\cr
&&&1&&&&\cr }\right),
$$

$$
\widetilde\Gamma_5=\left(\matrix{
&&&&&1&&\cr
&&&&1&&&\cr
&&&&&&&-1\cr
&&&&&&-1&\cr
&1&&&&&&\cr
1&&&&&&&\cr
&&&-1&&&&\cr
&&-1&&&&&\cr }\right),\quad
\widetilde\Gamma_6=\left(\matrix{
&&&&-1&&&\cr
&&&&&-1&&\cr
&&&&&&1&\cr
&&&&&&&1\cr
1&&&&&&&\cr
&1&&&&&&\cr
&&-1&&&&&\cr
&&&-1&&&&\cr }\right),
$$
$$
\widetilde\Gamma_7=\left(\matrix{
&&&&&1&&\cr
&&&&-1&&&\cr
&&&&&&&-1\cr
&&&&&&1&\cr
&1&&&&&&\cr
-1&&&&&&&\cr
&&&-1&&&&\cr
&&1&&&&&\cr }\right)\,.
$$

Since the spinor representation of $SO(5,2)$ is pseudoreal, we may
define symplectic Majorana spinors, $\chi^\alpha$, $\alpha=1,2$.  In
the present realization they satisfy
\eqn\appgsm{ \chi^\alpha=-i\epsilon^{\alpha\beta}\widetilde
\Gamma_2(\chi^\beta)^*\,,}
where $\epsilon^{12}=1$.

\goodbreak
\vskip2.cm\centerline{\bf Acknowledgements}
\noindent

We would like to thank A.~Hanany, P.~Mayr, J.~Polchinski, S.~Shenker,
A.~Strominger, and A.~Zaffaroni for perspicacious commentary. The research
of D.Z.F. was supported in part by the NSF under grant number
PHY-97-22072. The research of S.S.G.\ was supported by the Harvard Society
of Fellows, and also in part by the NSF under grant number PHY-98-02709,
and by DOE grant DE-FGO2-91ER40654.  S.S.G.\ also thanks the Institute for
Theoretical Physics at Santa Barbara, the University of Southern
California, Stanford University, and the University of Pennsylvania for
hospitality.  The work of K.P.\ and N.W.\ was supported in part by funds
provided by the DOE under grant number DE-FG03-84ER-40168.


\vskip 2cm

\listrefs

\vfill
\eject
\end